\documentclass[12pt]{article}

\usepackage{graphics}
\usepackage{amsmath}
\usepackage{epsfig}
\usepackage{cite}
\usepackage[english]{babel}

\usepackage[T2A]{fontenc} 
\usepackage{amssymb,bm,amsfonts,amsmath,textcomp,mathtext,indentfirst} 
\usepackage{misccorr,enumerate,cite,csquotes,tocloft,float,setspace,xcolor}
\usepackage{graphicx}
\usepackage[outdir=./]{epstopdf}
\graphicspath{{img/}}
\usepackage[singlelinecheck=false]{caption}
\title{Bottomonium production and polarization in the NRQCD with $k_T$-factorization. III: $\Upsilon(1S)$ and $\chi_b(1P)$ mesons}
\author{N.A.~Abdulov$^{1}$, A.V.~Lipatov$^{1,\,2}$}

\begin{document}

\maketitle

\begin{center}

{\it $^1$Skobeltsyn Institute of Nuclear Physics, Lomonosov Moscow State University, 119991 Moscow, Russia}\\
{\it $^2$Joint Institute for Nuclear Research, 141980 Dubna, Moscow Region, Russia}

\end{center} 

\vspace{0.1cm}

\begin{center}

{\bf Abstract }

\end{center} 

The $\Upsilon(1S)$ meson production and polarization at high energies is 
studied in the framework of the $k_T$-factorization approach.
Our consideration is based on the non-relativistic 
QCD formalism for a bound states formation and off-shell production
amplitudes for hard partonic 
subprocesses. The direct production mechanism, feed-down contributions 
from radiative $\chi_b(mP)$ decays and contributions from 
$\Upsilon(3S)$ and $\Upsilon(2S)$ decays are taken into account.
The transverse momentum dependent (TMD) gluon
densities in a proton were derived from the 
Ciafaloni-Catani-Fiorani-Marchesini
evolution equation and the Kimber-Martin-Ryskin prescription. 
Treating the non-perturbative color octet transitions 
in terms of multipole radiation theory, we extract 
the corresponding non-perturbative matrix elements  
for $\Upsilon(1S)$ and $\chi_b(1P)$ mesons
from a combined fit to transverse momenta distributions 
measured at various LHC experiments. 
Then we apply the extracted values 
to investigate the polarization 
parameters $\lambda_\theta$, $\lambda_\phi$ and $\lambda_{\theta\phi}$,
which determine the $\Upsilon(1S)$ spin density matrix. 
Our predictions
have a reasonably good agreement with the currently 
available Tevatron and LHC data within the theoretical and experimental uncertainties. 

\vspace{0.5cm}

\noindent{\it Keywords:} bottomonia, non-relativistic QCD, CCFM evolution, TMD gluon density

\newpage

Since it was first observed, the production of heavy quarkonium states 
in high energy hadronic collisions remains 
a subject of considerable theoretical and experimental interest\cite{lansberg,bramilla}. 
These processes are sensitive to the interaction dynamics both at small
and large distances: the production of heavy ($c$ or $b$) quarks with
a high transverse momentum is followed by a 
bound states formation with a low relative quark momentum. 
Accordingly, a theoretical description of these processes involves 
both perturbative and non-perturbative methods, as it was proposed 
in the non-relativistic QCD (NRQCD)\cite{bodwinBraaten,choLeibovich,kramer,maWang}.
However, it is known that the 
NRQCD at the next-to-leading order (NLO) accuracy 
meets difficulties in a simultaneous description of all the collider data in there entirety
(see also discussions\cite{lansbergShao,lansberg2,fengHe,gongLi,chaoMa,gongWan}).
In particular, it has a long-standing challenge in the $J/\psi$ and $\psi(2S)$
polarization and provides an inadequate description\cite{zhangSun, butenschonHe,likhoded2,hanMa,biswal,butenschonkniehl} of the recent $\eta_c$
production data taken by the LHCb Collaboration at the LHC\cite{LHCbEtac}.
One of possible solutions of the problems mentioned above, which implies a certain modification
of the NRQCD rules, has been proposed recently\cite{baranov}.
As it was shown, the approach\cite{baranov} allows one to 
describe well the recent data on the production and polarization of the
entire charmonia family.
The bottomonium production, namely $\Upsilon(nS)$ and $\chi_b(mP)$ mesons,
provides an alternative laboratory for understanding 
the physics of the hadronization of heavy quark pairs.
Due to heavier masses and a smaller quark relative velocity $v$ (in a produced quarkonium rest frame),
these processes could be even a more suitable case
to apply the double NRQCD expansion in QCD coupling $\alpha_s$ and $v$.
The NLO NRQCD predictions for the $\Upsilon(nS)$
production at the LHC were presented\cite{fengGong,hanMa2,chinese2021}.
Of course, it is important to apply also the approach\cite{baranov} to the bottomonium family.

Our present work continues the line started in the previous 
studies\cite{upsilonI,upsilonII}. We have considered there the inclusive 
production of $\Upsilon(3S)$, $\Upsilon(2S)$, $\chi_b(3P)$
and $\chi_b(2P)$ mesons and now come to $\Upsilon(1S)$ and $\chi_b(1P)$ mesons.
The motivation for the whole business has been already given\cite{upsilonI,upsilonII}.
Below we present a systematic analysis of
the CMS\cite{cms1,cms2,cmsb2b1}, ATLAS\cite{atlas} and LHCb\cite{lhcb1,lhcb2,lhcbr,lhcbb2b1} data
on the $\Upsilon(1S)$ and $\chi_b(1P)$ production
collected at $\sqrt s = 7$, $8$ and $13$~TeV (including the different 
relative production rates) and we extract from these data 
non-perturbative matrix elements (NMEs)
for the $\Upsilon(1S)$ and $\chi_b(1P)$ mesons.
Then we make predictions for polarization
parameters $\lambda_\theta$, $\lambda_\phi$, $\lambda_{\theta\phi}$ 
(and a frame-independent parameter $\tilde\lambda$), which determine 
the $\Upsilon(1S)$ spin density matrix and compare
them to the currently available data\cite{cmslam,cdf2}.
As it is known, the feed-down contributions from
$\chi_b(2P)$, $\chi_b(3P)$, $\Upsilon(2S)$ and $\Upsilon(3S)$
decays give a significant impact on the $\Upsilon(1S)$
production and polarization, so 
studies\cite{upsilonI,upsilonII} are important and
necessary for our present consideration.
Another important issue concerns the
relative production rate $\sigma(\chi_{b2})/\sigma(\chi_{b1})$
recently measured by the CMS\cite{cmsb2b1} and LHCb\cite{lhcbb2b1} Collaborations.
This ratio is sensitive to the color singlet (CS) and color octet (CO)
production mechanisms and provides information complementary to 
the study of the $S$-wave bottomonium states.

In the present note we follow mostly the same steps 
as in\cite{upsilonI,upsilonII}.
So, to describe the perturbative production of the $b\bar b$
pair in the hard scattering subprocesses we apply the
$k_T$-factorization approach\cite{sumkt1,sumkt2},
which is mainly based on the Balitsky-Fadin-Kuraev-Lipatov 
(BFKL)\cite{bfkl} or Ciafaloni-Catani-Fiorani-Marchesini (CCFM)\cite{ccfm} 
gluon evolution equations. A detailed description and
discussion of the different aspects of the $k_T$-factorization 
can be found in the reviews\cite{angeles}. As usual, we see certain advantages in
the ease of including into the calculations a large piece of higher order pQCD 
corrections taking them into account in the 
form of transverse momentum dependent (TMD), or unintegrated, gluon densities in a proton.
Our consideration is based on the off-shell gluon-gluon fusion 
subprocesses representing
the true leading order (LO) in QCD:
\begin{equation}
g^*(k_1) + g^*(k_2) \rightarrow \Upsilon[{}^3S_1^{(1)}](p) + g(k),
\end{equation}	
\begin{equation}
g^*(k_1) + g^*(k_2) \rightarrow \Upsilon[{}^1S_0^{(8)},{}^3S_1^{(8)},{}^3P_J^{(8)}](p).
\end{equation}	
\begin{equation}
g^*(k_1) + g^*(k_2) \rightarrow \chi_{bJ}(p)[{}^3P_J^{(1)},{}^3S_1^{(8)}] \rightarrow \Upsilon(p_1) + \gamma(p_2),
\end{equation}

\noindent where $J = 0, 1$ or $2$ and the four-momenta of all particles 
are given in the parentheses. The color states taken into account 
are directly indicated. Both initial gluons are off mass shell, that 
means that they have non-zero transverse four-momenta
$k_1^2 = - {\mathbf k}_{1T}^2 \neq 0$,
$k_2^2 = - {\mathbf k}_{2T}^2 \neq 0$
and an admixture of longitudinal component in the polarization four-vectors
(see\cite{sumkt1,sumkt2} for more information).
The corresponding off-shell ($k_T$-dependent)
production amplitudes contain projection operators\cite{colorcs} 
for spin and color, that guarantee the proper quantum 
numbers of the final state bottomonium.
Following the ideas\cite{baranov}, to describe the nonperturbative transformations 
of the color-octet $b\bar b$ pairs produced in hard subprocesses 
into observed final state mesons we employ the classical 
multipole radiation theory (where the electric dipole $E1$ 
transition dominates\cite{batuninCho})
under the key physical assumption 
that the lifetime of intermediate color-octet states is rather long.
According to\cite{baranov}, only a single $E1$ transition is needed to 
transform a $P$-wave state into an $S$-wave state\footnote{The corresponding $E1$ 
transition amplitudes are listed in\cite{batuninCho}.}, 
whereas the
transformation of the color-octet $S$-wave state into the color-singlet $S$-wave state 
is treated as two successive $E1$ transitions ${}^3S_1^{(8)} \rightarrow {}^3P_J^{(8)} + g$, 
${}^3P_J^{(8)} \rightarrow {}^3S_1^{(1)} + g$ proceeding via either of 
three intermediate ${}^3P_J^{(8)}$ states with $J = 0,1,2$.
An essential consequence of the idea above is the nonconservation of the spin momentum $S_z$
during the transformation of color octet $^3S_1^{(8)}$ state into the color-singlet $^3S_1^{(1)}$ one.
In fact, the intermediate $P$-wave state is a state with definite total momentum $J$ and 
its projection $J_z$ rather than a state with definite $L_z$ and $S_z$.
To describe the formation of the intermediate state, we have to contract
the electric dipole transition amplitude\cite{batuninCho} (which by its own conserves $S_z$)
with Clebsch-Gordan coefficients $|L,\,S,\,J,\,J_z\rangle\langle L,\,L_z,\,S,\,S_z|$
which are symmetric with respect to $L_z$ and $S_z$.
Then the resulting expression comprises both: the terms containing
$\epsilon^*(S_z[^3S_1])\cdot\epsilon(S_z[^3P_J])$ and the terms containing
$\epsilon^*(S_z[^3S_1])\cdot\epsilon(L_z[^3P_J])$.
Consequently, there is no direct transfer from the initial spin polarization 
to the final spin polarization (see\cite{baranov} for more information).

Below we apply the gauge invariant expressions 
for quarkonia production and decay amplitudes implemented into 
the Monte-Carlo event generator \textsc{pegasus}\cite{pegasus}. 
The derivation steps are explained in~\cite{upsilonI,upsilonII} in detail.

According to the $k_T$-factorization prescription, to calculate the cross 
sections of a considered process one has to convolute the 
partonic cross section $\hat \sigma^*$ (related with an off-shell production 
amplitude) and TMD gluon densities in a proton $f_g(x, {\mathbf k}_{T}^2, \mu^2)$:
\begin{equation}
  \sigma = \int dx_1 dx_2 d{\mathbf k}^2_{1T} d{\mathbf k}^2_{2T} \, \hat \sigma^*(x_1, x_2, {\mathbf k}_{1T}^2, {\mathbf k}_{2T}^2, \mu^2) f_g(x_1, {\mathbf k}_{1T}^2, \mu^2) f_g(x_2, {\mathbf k}_{2T}^2, \mu^2){{d\phi_1}\over{2\pi}}{{d\phi_2}\over{2\pi}}, 
\label{x-section}
\end{equation}

\noindent 
where $x_1$ and $x_2$ are the longitudinal momentum fractions of initial off-shell gluons, $\phi_1$ 
and $\phi_2$ are their azimuthal angles and $\mu$ is the hard interaction scale.
Following\cite{upsilonI,upsilonII}, we have tested several sets of TMD gluon densities in a proton. 
Two of them (A0\cite{A0set} and JH'2013~set~1\cite{jhs}) were obtained from the CCFM equation 
where all input parameters were fitted to the proton structure function $F_2(x,Q^2)$. 
We have applied the TMD gluon densities obtained within the Kimber-Martin-Ryskin (KMR) prescription\cite{KMR},
which provides a method to construct the TMD quark and gluon distributions 
from the conventional (collinear) ones. 
For the input, we have applied the recent LO NNPDF3.1 set\cite{nnpdf}. 
The parton level calculations according to (\ref{x-section})
were performed using the Monte-Carlo generator \textsc{pegasus}. 	
Of course, we take into account the feed-down 
contributions from $\chi_b(3P)$, $\chi_b(2P)$, $\chi_b(1P)$, 
$\Upsilon(3S)$ and $\Upsilon(2S)$ decays.

Numerically, everywhere we set the masses  
$m_{\Upsilon(1S)} = 9.4603$~GeV, $m_{\Upsilon(2S)} = 10.02326$~GeV, $m_{\Upsilon(3S)} = 10.3552$~GeV,
$m_{\chi_{b1}(3P)} = 10.512$~GeV, $m_{\chi_{b2}(3P)} = 10.522$~GeV, 
$m_{\chi_{b0}(2P)} = 10.232$~GeV, $m_{\chi_{b1}(2P)} = 10.255$~GeV, $m_{\chi_{b2}(2P)} = 10.268$~GeV,
$m_{\chi_{b0}(1P)} = 9.8594$~GeV, $m_{\chi_{b1}(1P)} = 9.8928$~GeV, $m_{\chi_{b2}(1P)} = 9.9122$~GeV\cite{pdg} 
and adopt the usual non-relativistic approximation $m_b = m_{\cal Q}/2$
for the beauty quark mass, where $m_{\cal Q}$ is the mass of 
bottomonium $\cal Q$.
We set the necessary branching ratios as they are given in\cite{pdg}. 
Note that there is no experimental data for the branching 
ratios of $\chi_b(3P)$,
so we use the results of assumption\cite{hanMa2}
that the total decay widths of $\chi_b(mP)$ are approximately 
independent on $m$. So, we have
$B(\chi_{b1}(3P)\rightarrow \Upsilon(1S) + \gamma) = 0.0381$ and
$B(\chi_{b2}(3P) \rightarrow \Upsilon(1S) + \gamma) = 0.0192$\cite{hanMa2}.
We use the one-loop formula for the QCD coupling $\alpha_s$ 
with $n_f = 4(5)$ quark flavours at $\Lambda_{\rm QCD} = 250(167)$~MeV 
for A0 (KMR) gluon density
and two-loop expression for $\alpha_s$ with $n_f = 4$ 
and $\Lambda_{\rm QCD} = 200$~MeV for JH'2013 set 1.   
We set the color-singlet NMEs $\langle\mathcal{O}^{\Upsilon(1S)}[{}^{3}S_1^{(1)}]\rangle  = 8.39$~GeV$^3$ 
and $\langle\mathcal{O}^{\chi_{b0}(1P)}[{}^{3}P_0^{(1)}]\rangle  = 2.30$~GeV$^5$ 
as obtained from the potential model calculations~\cite{eichten}.
All the NMEs for $\Upsilon(2S)$, $\Upsilon(3S)$, $\chi_b(2P)$ and $\chi_b(3P)$ mesons were derived in\cite{upsilonI,upsilonII}.

To determine the NMEs for 
both $\Upsilon(1S)$ and $\chi_b(1P)$ mesons
we have performed a global fit to the $\Upsilon(1S)$ production data at 
the LHC.
We have included in the fitting procedure the $\Upsilon(1S)$ transverse 
momentum distributions measured by
the CMS~\cite{cms1,cms2} and ATLAS~\cite{atlas} Collaborations 
at $\sqrt s = 7$ and $13$~TeV.
Similar to the NRQCD analyses\cite{fengGong,hanMa2,chinese2021}, we have excluded from our fit the low $p_T$ region and considered 
only data at
$p_T > p_T^{\rm cut} = 10$~GeV. 
We note that at low transverse momenta
a more accurate treatment of large logarithms $\sim\ln m^2_{\Upsilon}/p^2_T$ and other nonperturbative effects become necessary.
To determine NMEs for $\chi_b(1P)$ mesons, we also included into the fit 
the recent LHCb data~\cite{lhcbr} on the radiative 
$\chi_b(1P) \to \Upsilon(1S) + \gamma$ decays 
collected at $\sqrt s = 7$ and $8$~TeV and 
the recent CMS~\cite{cmsb2b1} and LHCb data~\cite{lhcbb2b1} on the ratio
$\sigma(\chi_{b2}(1P))/\sigma(\chi_{b1}(1P))$ 
collected at $\sqrt s = 8$~TeV. 

Our analysis strategy is the following.
First,
we found that the $p_T$ shape of the direct $\Upsilon[^3S_1^{(8)}]$ and 
feed-down $\chi_b[^3S_1^{(8)}]$ contributions to the $\Upsilon(1S)$ production is almost 
the same in all kinematical regions
probed at the LHC. Thus, the ratio
\begin{equation}
r = { \sum\limits_{J = 0}^{2} (2J+1) \, B(\chi_{bJ}(1P) \to \Upsilon(1S) + \gamma) d\sigma[\chi_{bJ}(1P), {}^3S_1^{(8)}]/dp_T \over d\sigma [\Upsilon(1S), {}^3S_1^{(8)}]/dp_T }
\label{eqr}
\end{equation}

\noindent 
can be well approximated by a constant for a wide $\Upsilon(1S)$ 
transverse momentum $p_T$ and rapidity $y$ range at different energies. 
For example, we estimate the mean-square average $r = 1.743 \pm 0.010$ for the A0 set, which is 
practically the same for all other TMD gluon densities in a proton.
So, we construct a linear combination
\begin{equation}
M_r = \langle\mathcal{O}^{\Upsilon(1S)}[{}^{3}S_1^{(8)}]\rangle + r \langle\mathcal{O}^{\chi_{b0}(1P)}[{}^{3}S_1^{(8)}]\rangle,
\label{eqMr}
\end{equation}
\noindent
which can be only extracted from the measured $\Upsilon(1S)$ transverse
momentum distributions.
Note that here we considered the color singlet
wave functions of $\chi_b(1P)$ mesons as independent (not necessarily identical) free parameters,
as it was proposed\cite{baranovc2c1} to describe the LHC data on  
relative $\sigma(\chi_{c2})/\sigma(\chi_{c1})$ production rate.
Of course, we understand that doing so is at odds with the Heavy Quark Effective Theory
(HQET) and Heavy Quark Spin Symmetry (HQSS).
However, it was argued\cite{baranovc2c1} that the HQSS predictions must not be taken for 
granted\footnote{The possible reason may be seen in the
spin-orbilal interactions or in radiative corrections which 
can be large (see more discussion\cite{chicJrecent}).}.
Thus, here we try two alternative scenarios for the mesons. 
We assume HQSS violation either solely for the 
color singlet states ("fit A") or for both color singlet and color octet states 
("fit B"). In the latter case, the color octet NMEs for $\chi_b(1P)$ mesons 
are also treated as independent 
parameters not related to each other through the $(2J+1)$ factor.
Thus, we introduce the ratios:
\begin{equation}
r^{CO}_0 = { B(\chi_{b0}(1P) \to \Upsilon(1S) + \gamma) d\sigma[\chi_{b0}(1P), {}^3S_1^{(8)}]/dp_T \over d\sigma [\Upsilon(1S), {}^3S_1^{(8)}]/dp_T },
\label{eqrco0}
\end{equation}
\begin{equation}
r^{CO}_1 = { B(\chi_{b1}(1P) \to \Upsilon(1S) + \gamma) d\sigma[\chi_{b1}(1P), {}^3S_1^{(8)}]/dp_T \over d\sigma [\Upsilon(1S), {}^3S_1^{(8)}]/dp_T },
\label{eqrco1}
\end{equation}
\begin{equation}
r^{CO}_2 = { B(\chi_{b2}(1P) \to \Upsilon(1S) + \gamma) d\sigma[\chi_{b2}(1P), {}^3S_1^{(8)}]/dp_T \over d\sigma [\Upsilon(1S), {}^3S_1^{(8)}]/dp_T },
\label{eqrco2}
\end{equation}
\noindent
and obtain the mean-square average values 
$r^{CO}_0 = 0.01737 \pm 0.00006$, $r^{CO}_1 = 0.304 \pm 0.002$ and $r^{CO}_2 = 0.1626 \pm 0.0010$ (for the A0 gluon density). 
Then, instead of (\ref{eqMr}), we have a modified linear combination for the color octet NMEs:
\begin{equation}
M_{r} = \langle\mathcal{O}^{\Upsilon(1S)}[{}^{3}S_1^{(8)}]\rangle + \sum\limits_{J = 0}^{2} (2J+1) \,  r^{CO}_J \langle\mathcal{O}^{\chi_{bJ}(1P)}[{}^{3}S_1^{(8)}]\rangle.
\label{eqMrMod}
\end{equation}
\noindent
Next, we found that the $p_T$ shapes of the direct $\Upsilon[^3P_J^{(8)}]$,  
feed-down $\chi_b[^3P_1^{(1)}]$ and $\chi_b[^3P_2^{(1)}]$ contributions to the $\Upsilon(1S)$ production are also
the same in all kinematical regions. So, the ratios
\begin{equation}
r_1 = { B(\chi_{b2}(1P) \to \Upsilon(1S) + \gamma) d\sigma[\chi_{b2}(1P), {}^3P_2^{(1)}]/dp_T \over B(\chi_{b1}(1P) \to \Upsilon(1S) + \gamma) d\sigma[\chi_{b1}(1P), {}^3P_1^{(1)}]/dp_T },
\label{eqr1}
\end{equation}
\begin{equation}
r_2 = { \sum\limits_{J = 0}^{2} (2J+1) \, d\sigma [\Upsilon(1S), {}^3P_J^{(8)}]/dp_T \over B(\chi_{b1}(1P) \to \Upsilon(1S) + \gamma) d\sigma[\chi_{b1}(1P), {}^3P_1^{(1)}]/dp_T }
\label{eqr2}
\end{equation}

\noindent 
can be approximated by constants for a wide $\Upsilon(1S)$ 
transverse momentum $p_T$ and rapidity $y$ range at different energies. 
For example, we estimate the mean-square average $r_1 = 0.91 \pm 0.02$ and $r_2 = 104 \pm 2$ for the A0 set.
Then we construct a linear combination
\begin{equation}
M_{r_1r_2} = \langle\mathcal{O}^{\chi_{b1}(1P)}[{}^{3}P_1^{(1)}]\rangle + r_1 \langle\mathcal{O}^{\chi_{b2}(1P)}[{}^{3}P_2^{(1)}]\rangle + r_2 \langle\mathcal{O}^{\Upsilon(1S)}[{}^{3}P_0^{(8)}]\rangle,
\end{equation}
\noindent
which can be extracted from the measured $\Upsilon(1S)$ transverse
momentum distributions.
As the next step, we use the recent LHCb data~\cite{lhcbr} on the ratio
of $\Upsilon(1S)$ mesons 
originating from the $\chi_b(1P)$ radiative decays measured
at $\sqrt s = 7$ and $8$~TeV:
\begin{equation}
R^{\chi_b(1P)}_{\Upsilon(1S)} = \sum\limits_{J = 1}^{2} {\sigma(pp\rightarrow \chi_{bJ}(1P) + X) \over \sigma(pp \rightarrow \Upsilon(1S) + X)} \times B(\chi_{bJ} \to \Upsilon(1S) + \gamma).
\label{eqrUp}
\end{equation}

\noindent
In the "fit A" scenario, from the known $M_r$,  $M_{r_1r_2}$ and $R^{\chi_b(1P)}_{\Upsilon(1S)}$ values
one can separately determine the 
$\langle\mathcal{O}^{\Upsilon(1S)}[{}^{3}S_1^{(8)}]\rangle$,
$\langle\mathcal{O}^{\chi_{b0}(1P)}[{}^{3}S_1^{(8)}]\rangle$,
$\langle\mathcal{O}^{\Upsilon(1S)}[{}^{3}P_0^{(8)}]\rangle$ and 
the linear combination $M_{CS} = \langle\mathcal{O}^{\chi_{b1}(1P)}[{}^{3}P_1^{(1)}]\rangle + r_1 \langle\mathcal{O}^{\chi_{b2}(1P)}[{}^{3}P_2^{(1)}]\rangle$.
In the case of "fit B", we can determine the 
$\langle\mathcal{O}^{\Upsilon(1S)}[{}^{3}S_1^{(8)}]\rangle$,
$\langle\mathcal{O}^{\chi_{b0}(1P)}[{}^{3}S_1^{(8)}]\rangle$,
$\langle\mathcal{O}^{\Upsilon(1S)}[{}^{3}P_0^{(8)}]\rangle$ and two linear combinations
$M_{CS} = \langle\mathcal{O}^{\chi_{b1}(1P)}[{}^{3}P_1^{(1)}]\rangle + r_1 \langle\mathcal{O}^{\chi_{b2}(1P)}[{}^{3}P_2^{(1)}]\rangle$ and 
$M_{CO} = \langle\mathcal{O}^{\chi_{b1}(1P)}[{}^{3}S_1^{(8)}]\rangle + r^{CO}_2/r^{CO}_1 \langle\mathcal{O}^{\chi_{b2}(1P)}[{}^{3}S_1^{(8)}]\rangle$.
Finally, we use recent CMS~\cite{cmsb2b1} and LHCb data~\cite{lhcbb2b1} measured
at $\sqrt s = 8$~TeV on the ratio
\begin{equation}
R^{\chi_{b2}(1P)}_{\chi_{b1}(1P)} = {\sigma(\chi_{b2}(1P)) \over \sigma(\chi_{b1}(1P))}.
\label{eqrb12}
\end{equation}

\noindent
From the known $M_{CS}$, $\langle\mathcal{O}^{\chi_{b0}(1P)}[{}^{3}S_1^{(8)}]\rangle$ and 
$R^{\chi_{b2}(1P)}_{\chi_{b1}(1P)}$ values
one can separately determine the $\langle\mathcal{O}^{\chi_{b1}(1P)}[{}^{3}P_1^{(1)}]\rangle$ and 
$\langle\mathcal{O}^{\chi_{b2}(1P)}[{}^{3}P_2^{(1)}]\rangle$ values for the first fit.
For the second one we use only the CMS data\cite{cmsb2b1}, because the LHCb data\cite{lhcbb2b1} 
are very few
and only increase the total error of the fitted quantities. 
So, from the known $M_{CS}$, $M_{CO}$ and 
$R^{\chi_{b2}(1P)}_{\chi_{b1}(1P)}$ we determine the $\langle\mathcal{O}^{\chi_{b1}(1P)}[{}^{3}P_1^{(1)}]\rangle$, 
$\langle\mathcal{O}^{\chi_{b2}(1P)}[{}^{3}P_2^{(1)}]\rangle$, $\langle\mathcal{O}^{\chi_{b1}(1P)}[{}^{3}S_1^{(8)}]\rangle$,
$\langle\mathcal{O}^{\chi_{b2}(1P)}[{}^{3}S_1^{(8)}]\rangle$ values. Therefore, we have
reconstructed the full map of the NMEs for both $\Upsilon(1S)$ and $\chi_b(1P)$ mesons. 

The fitting procedure described above was separately done in each of the rapidity 
subdivisions (using the fitting algorithm as implemented 
in the commonly used \textsc{gnuplot} package~\cite{gnuplot}) 
under the requirement that all the NMEs are strictly positive. 
Then, the mean-square average of
the fitted values was taken. The corresponding uncertainties are 
estimated in the conventional way
using Student's t-distribution at the confidence level $P = 80$\%.
The results of our fits are collected in Tables~\ref{tab1} and \ref{tab1.5}. For comparison, 
we also presented 
there the NMEs obtained in the conventional NLO NRQCD by other 
authors~\cite{fengGong}.
Note that the results\cite{fengGong} were obtained from the fit on the same 
data set as in our analysis.
The corresponding $\chi^2/d.o.f.$ 
are listed in Table~\ref{tab2}, where we additionally
show their dependence on the minimal $\Upsilon(1S)$ transverse momenta 
involved into the fit $p_T^{\rm cut}$.
As one can see, the $\chi^2/d.o.f.$ tends to stay the same or slightly increase when 
$p_T^{\rm cut}$ grows up and the
best fit of the LHC data is achieved with the A0 and KMR gluon, although 
other gluon densities also return reliable $\chi^2/d.o.f.$ values.
We note that including into the fit the latest CMS data~\cite{cms2} taken at $\sqrt s = 13$~TeV leads to
2 --- 3 times higher values of $\chi^2/d.o.f.$, as it was with the data on $\Upsilon(2S)$ \cite{upsilonII}. We have checked that this is true for both the $k_T$-factorization 
and collinear approaches\footnote{We have used the on-shell production amplitudes for color-octet $2 \to 2$ 
subprocesses from~\cite{choLeibovich}.} and, therefore, it could be a sign of some inconsistency 
between these CMS data and other measurements.

Both fit scenarios result in unequal 
values for $\chi_{b1}(1P)$ and $\chi_{b2}(1P)$ color singlet wave functions.
So, for "fit A" we achieved the ratio $\langle\mathcal{O}^{\chi_{b2}(1P)}[{}^{3}P_2^{(1)}]\rangle : 
\langle\mathcal{O}^{\chi_{b1}(1P)}[{}^{3}P_1^{(1)}]\rangle : 
\langle\mathcal{O}^{\chi_{b0}(1P)}[{}^{3}P_0^{(1)}]\rangle \sim 2.6 : 4.8 : 1$ for the JH'2013 set 1,  
$\sim 2.6 : 3.9 : 1$ for the KMR and 
$\sim 1 : 3 : 1$ for the A0 gluon densities, respectively.
This is an obvious contradiction with naive expectations based on the number
of spin degrees of freedom, $\sim 5 : 3 : 1$.
The difference between the predictions for this ratio
obtained with the considered TMD gluon densities 
could be a sign of a sensitivity 
of the relative production rate $R^{\chi_{b2}(1P)}_{\chi_{b1}(1P)}$ to the gluon distributions and/or 
due to lack of the experimental data.
If we assume the HQSS violation in the color octet sector
as well (the "fit B" scenario), the fitted values of the color singlet NMEs of the $\chi_{b1}(1P)$ and $\chi_{b2}(1P)$
mesons also differ from each other (see Table\ref{tab1.5}).
The latter qualitatively agrees with the observations~\cite{baranovc2c1, chicJrecent}
done in the case of the $\chi_c$ mesons.

\begin{figure}
\begin{center}
\includegraphics[width=7.0cm]{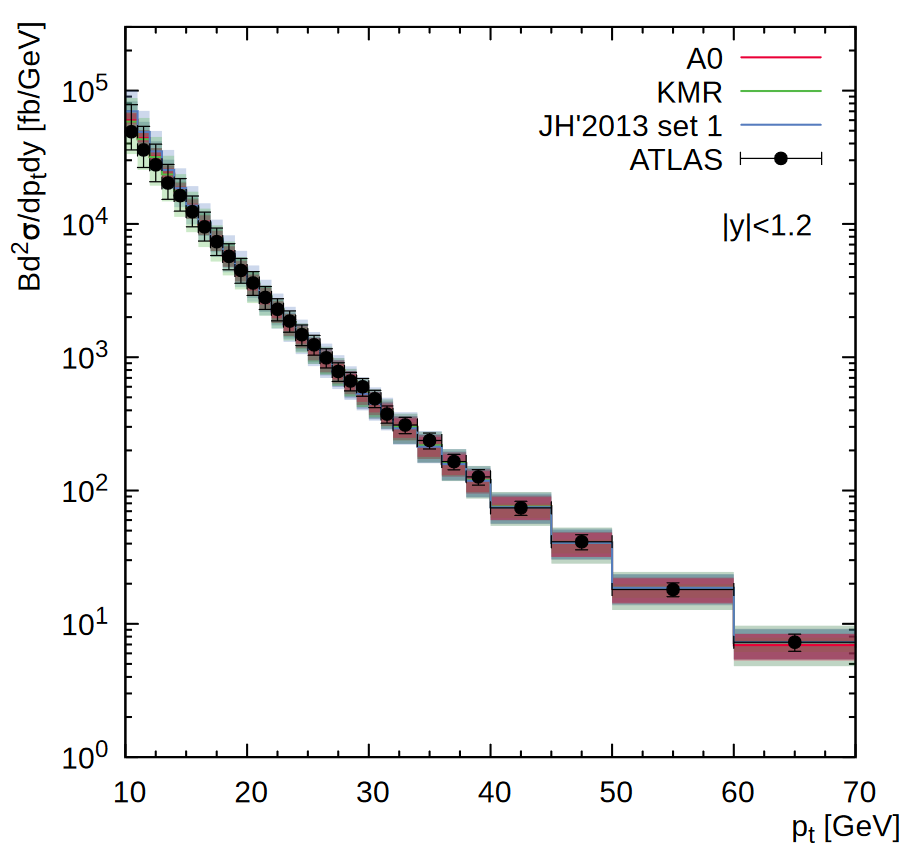}
\includegraphics[width=7.0cm]{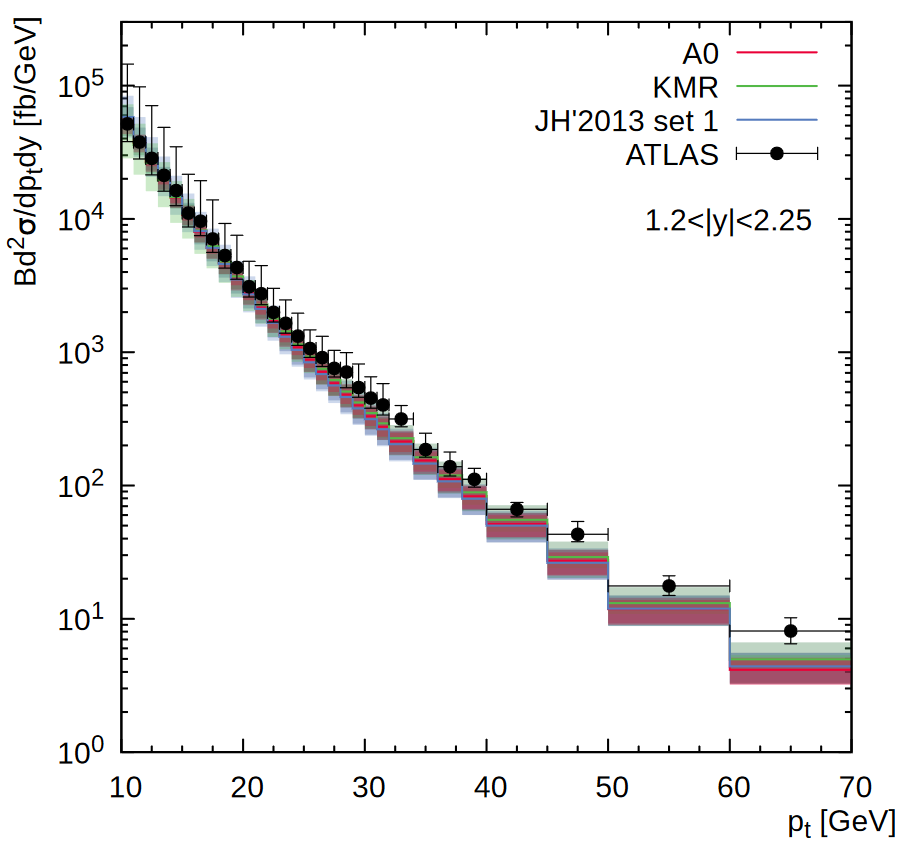}
\caption{Transverse momentum distribution of the 
  inclusive $\Upsilon(1S)$ production calculated at 
  $\sqrt s = 7$~TeV in the different rapidity regions. 
  The red, green and blue histograms
  correspond to the predictions obtained with the A0, KMR and JH'2013 set 1
  gluon densities. Shaded bands represent the total uncertainties 
  of our calculations, as it is described in the text.
  The experimental data are from ATLAS\cite{atlas}.}
\label{fig1}
\end{center}
\end{figure}

\begin{figure}
\begin{center}
\includegraphics[width=7.0cm]{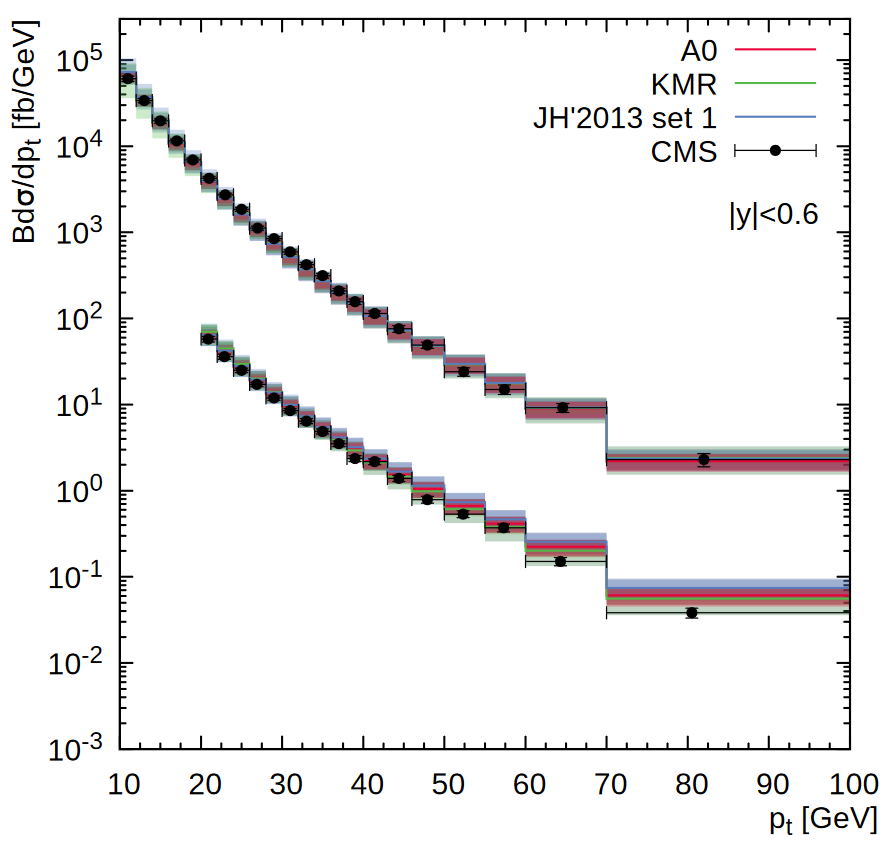}
\includegraphics[width=7.0cm]{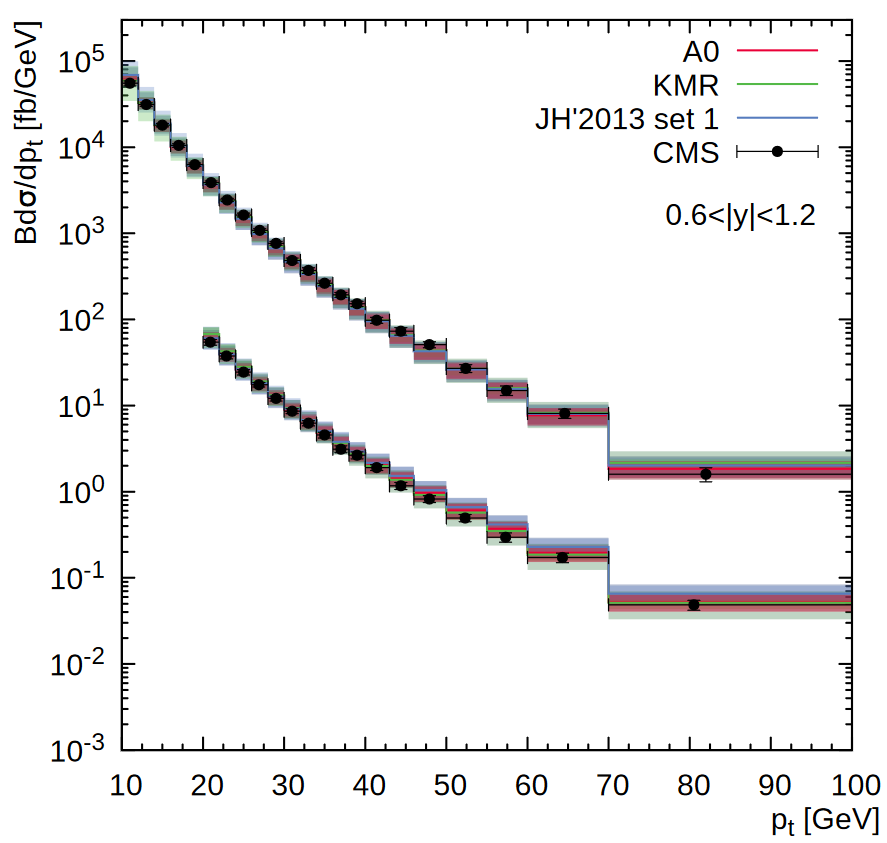}
\includegraphics[width=7.0cm]{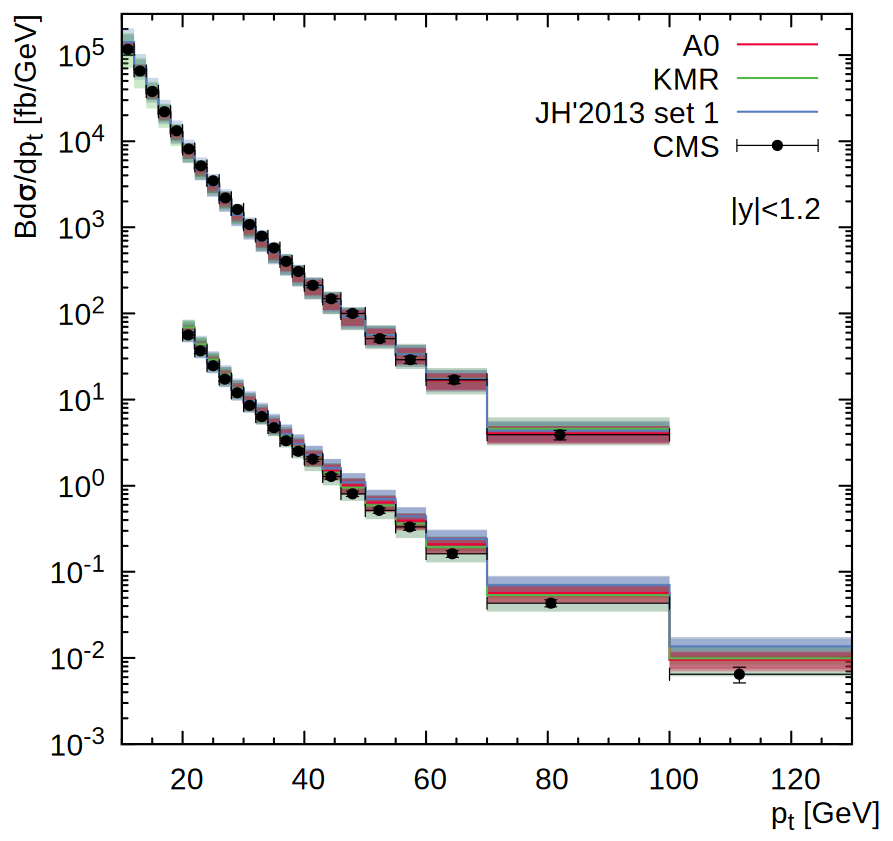}
\caption{Transverse momentum distribution of the
  inclusive $\Upsilon(1S)$ production calculated at $\sqrt s = 7$~TeV
  (upper histograms) and $\sqrt s = 13$~TeV (lower histograms, 
  divided by $100$) in the different rapidity regions. 
  The notation of all histograms is the same as in Fig.~\ref{fig1}.
  The experimental data are from CMS \cite{cms1,cms2}.}
\label{fig2}
\end{center}
\end{figure}

\begin{figure}
\begin{center}
\includegraphics[width=7.0cm]{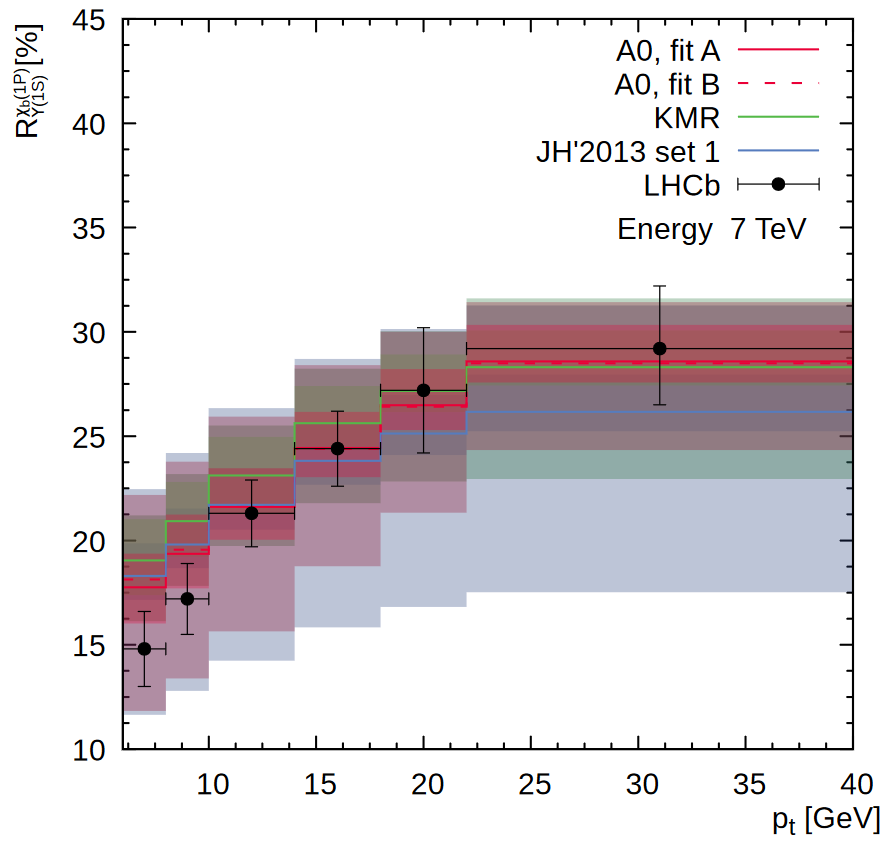}
\includegraphics[width=7.0cm]{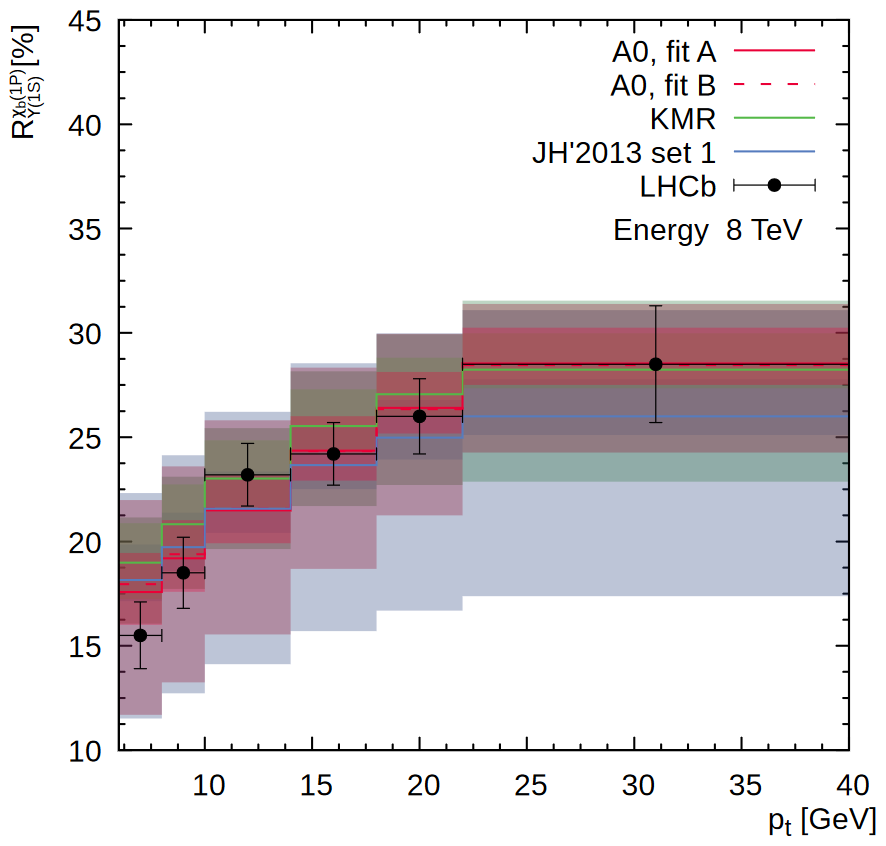}
\includegraphics[width=7.0cm]{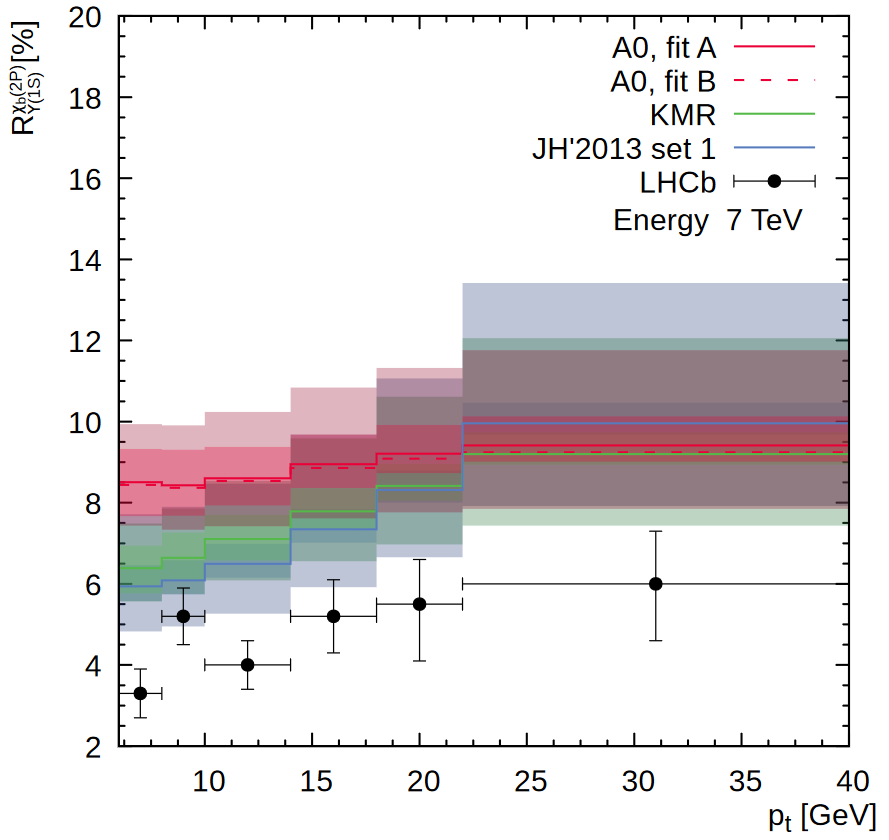}
\includegraphics[width=7.0cm]{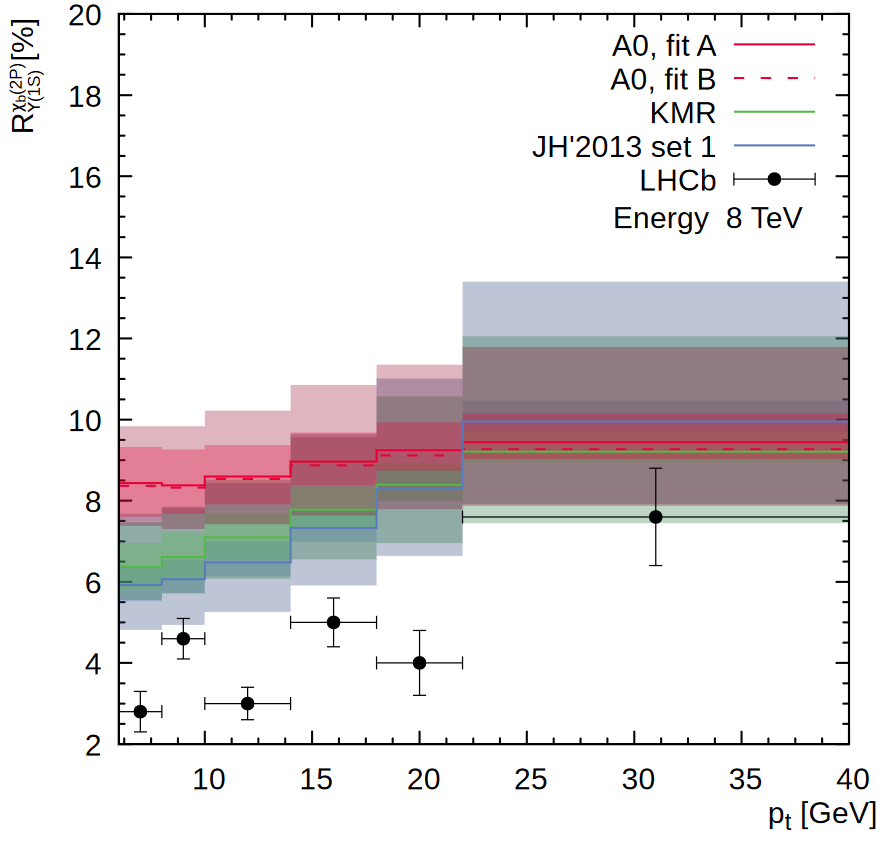}
\includegraphics[width=7.0cm]{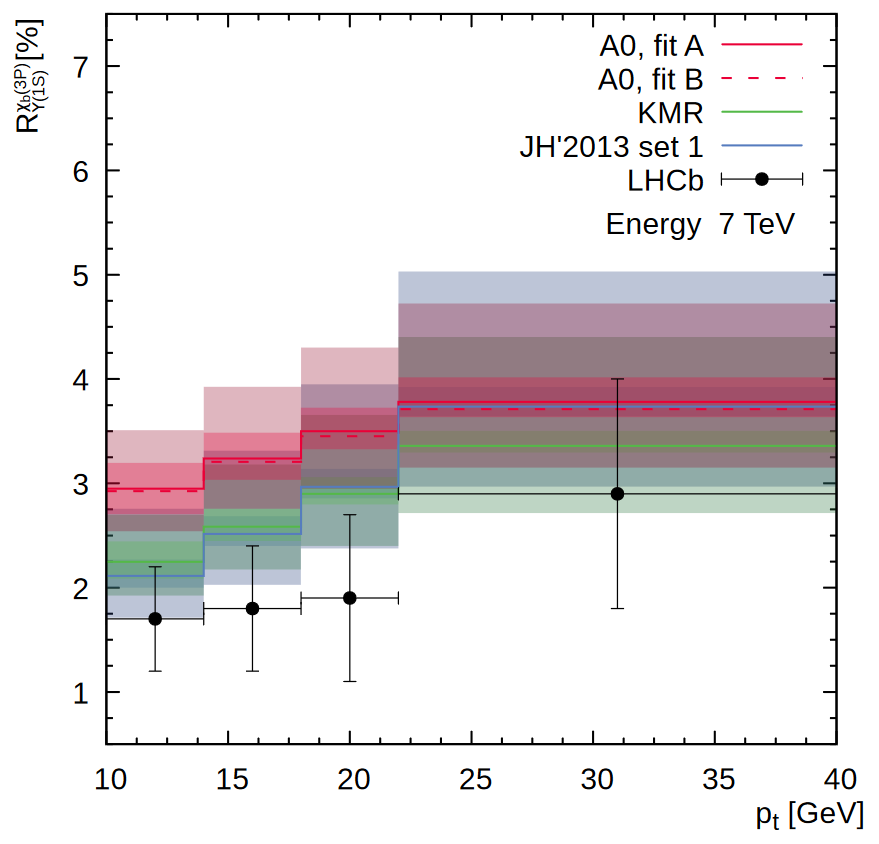}
\includegraphics[width=7.0cm]{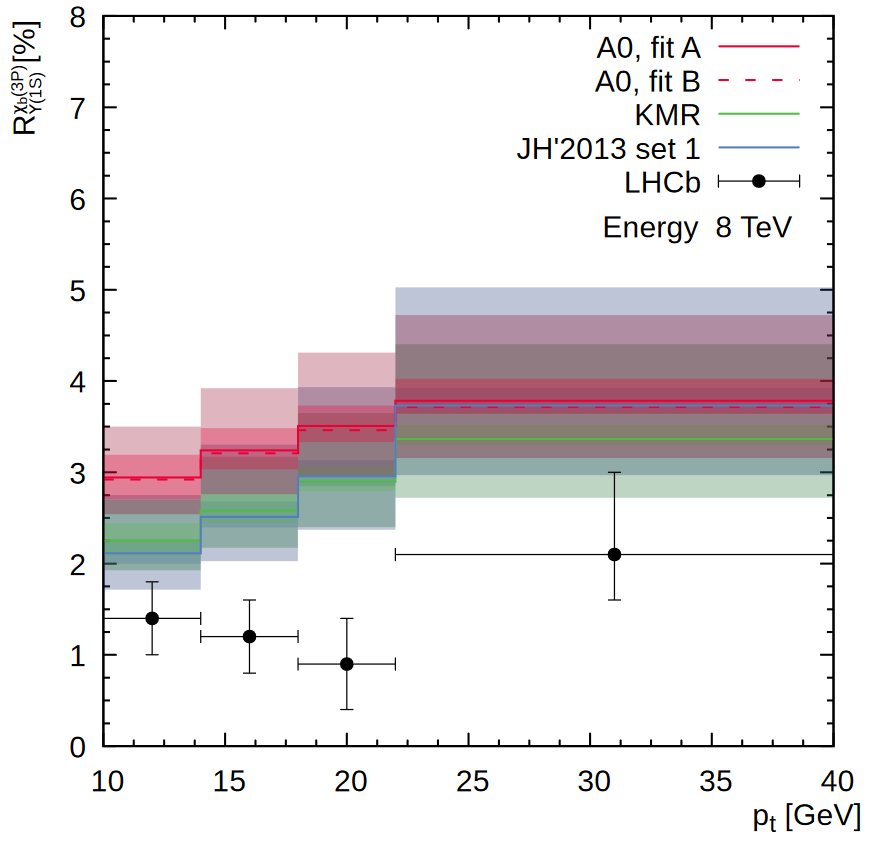}
\caption{The ratio $R^{\chi_b(mP)}_{\Upsilon(1S)}$
  calculated as a function of the $\Upsilon(1S)$ transverse momentum 
  calculated at $\sqrt s = 7$ and $8$~TeV using the NMEs from Tables 1 and 2 (for the A0 gluon density).
  The notation of all histograms is the same as in Fig.~\ref{fig1}.
  The experimental data are from LHCb \cite{lhcbr}.}
\label{fig3}
\end{center}
\end{figure}

\begin{figure}
\begin{center}
\includegraphics[width=7.0cm]{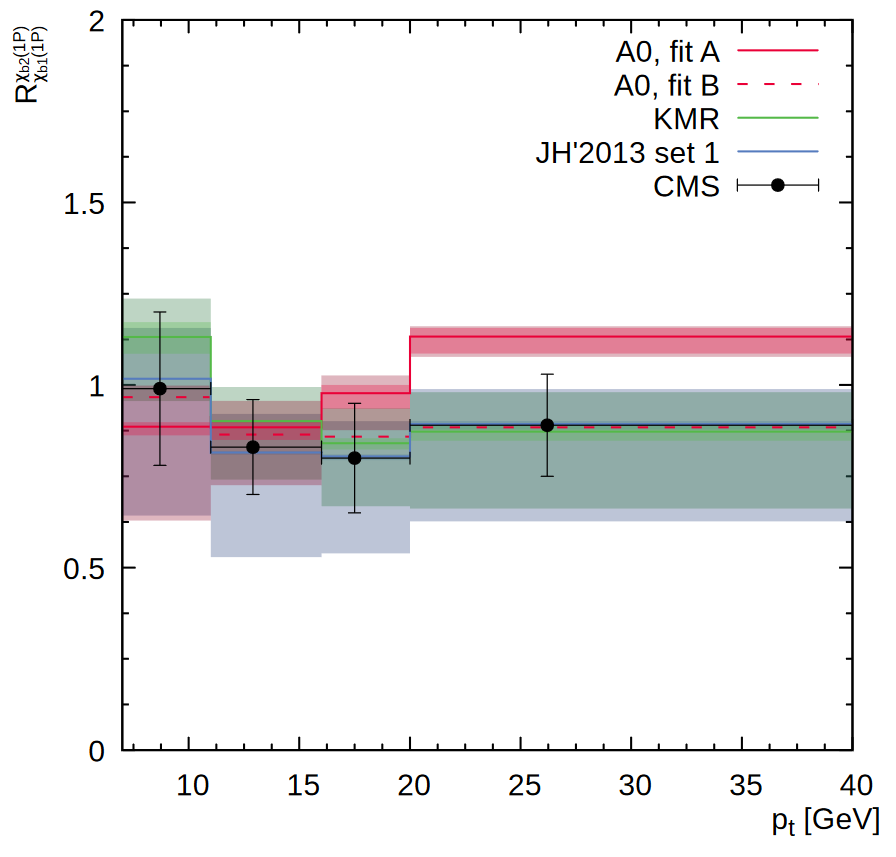}
\includegraphics[width=7.15cm]{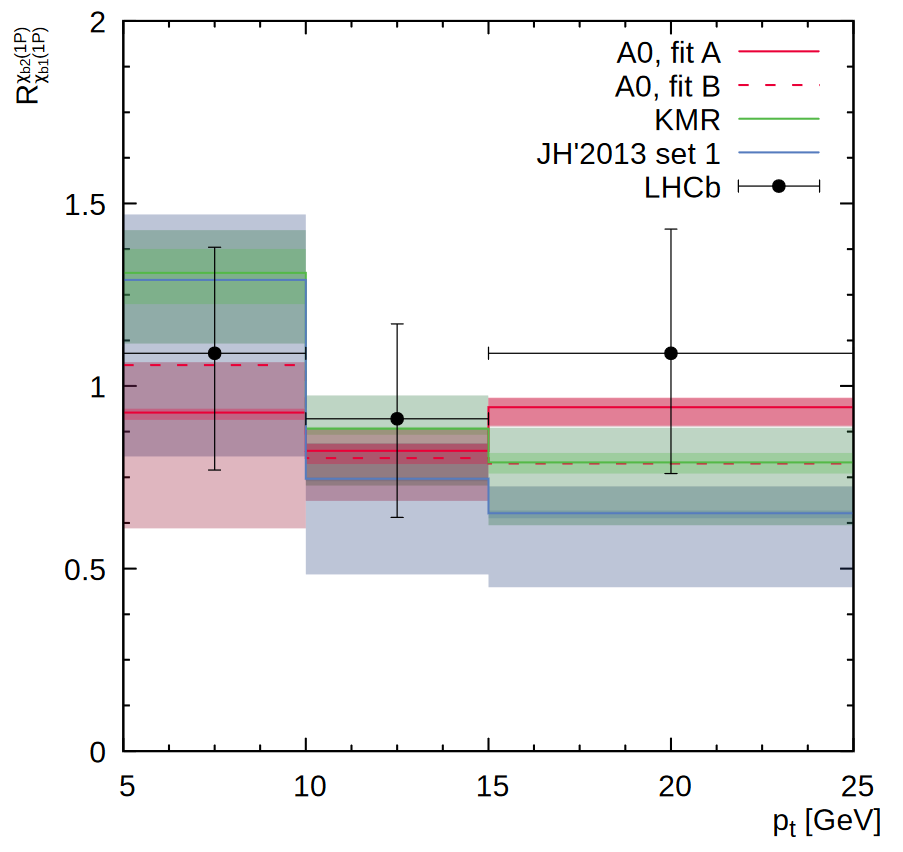}
\caption{The ratio $R^{\chi_{b2}(1P)}_{\chi_{b1}(1P)}$
  calculated as a function of the $\Upsilon(1S)$ transverse momentum 
  calculated at $\sqrt s = 8$~TeV using the NMEs from Tables 1 and 2 (for the A0 gluon density).
  The notation of all histograms is the same as in Fig.~\ref{fig1}.
  The experimental data are from CMS\cite{cmsb2b1} and LHCb\cite{lhcbb2b1}.}
\label{fig4}
\end{center}
\end{figure}

All the data used in the fits above are compared with our predictions 
in Figs.~\ref{fig1} --- \ref{fig4}.
The shaded areas represent the theoretical uncertainties of our 
calculations, 
which include the uncertainties coming from the NME 
fitting procedure and the scale uncertainties. 
To estimate the latter,
the standard variations in 
default renormalization scale (which is set to be equal to 
$\mu_R^2 = m_{\cal Q}^2 + {\mathbf p}_T^2$),
namely, $\mu_R\to 2\mu_R$ or $\mu_R\to\mu_R/2$ were introduced 
with replacing the A0 and JH'2013 set 1 gluon densities
by the A0$+$ and JH'2013 set 1$+$, or 
by the A0$-$ and JH'2013 set 1$-$ ones.
This was done to preserve the intrinsic correspondence between 
the TMD gluon set and the factorization scale taken as 
$\mu_F^2 = \hat s + {\mathbf Q}_T^2$ (where ${\mathbf Q}_T$ is the net 
transverse momentum of incoming off-shell gluon pair) according to the TMD 
gluon fits (see~\cite{A0set,jhs} for more information).
Of course, in the case of KMR gluons both 
factorization and renormalization scales have been varied to 
estimate the scale uncertainties.
One can see that we have achieved a reasonably good 
description of the CMS~\cite{cms1,cms2} and ATLAS~\cite{atlas} 
data for the $\Upsilon(1S)$ transverse 
momentum distributions in the whole $p_T$ range within the experimental and 
theoretical uncertainties. 
The relative production rates $R^{\chi_{b2}(1P)}_{\chi_{b1}(1P)}$
measured by the CMS\cite{cmsb2b1} and LHCb\cite{lhcbb2b1}
Collaborations and the $R^{\chi_b(1P)}_{\Upsilon(1S)}$
ratios measured by the LHCb Collaboration~\cite{lhcbr}
at $\sqrt s = 7$ and $8$~TeV are also reproduced well.
However, our predictions for the
$R^{\chi_b(2P)}_{\Upsilon(1S)}$ and $R^{\chi_b(3P)}_{\Upsilon(1S)}$ rates
tend to overestimate a bit the LHCb data~\cite{lhcbr},
although they are rather close to the measurements
within the uncertainties bands (see Fig.~\ref{fig3}).
The same situation is observed in the conventional 
NRQCD scenario, where the NLO NRQCD calculations\cite{fengGong,hanMa2,chinese2021} also overestimate
the experimental data for the $R^{\chi_b(2P)}_{\Upsilon(1S)}$ and 
$R^{\chi_b(3P)}_{\Upsilon(1S)}$ rates.
Evaluation of these observables involves the NMEs for 
$\Upsilon(2S)$, $\Upsilon(3S)$, $\chi_b(2P)$ and $\chi_b(3P)$
mesons determined previously\cite{upsilonI,upsilonII}.
Of course, both scenarios, "fit A" and "fit B", lead to exactly the 
same results for the $Y(1S)$ transverse momentum distributions
due to full correspondence between~(\ref{eqMr}) and (\ref{eqMrMod}).
The corresponding predictions differ to each other for the relative production
rates $R^{\chi_b(mP)}_{\Upsilon(nS)}$ and/or $R^{\chi_{b2}(1P)}_{\chi_{b1}(1P)}$ only (see Figs.~\ref{fig3} and~\ref{fig4}).

\begin{figure}
\begin{center}
\includegraphics[width=7.0cm]{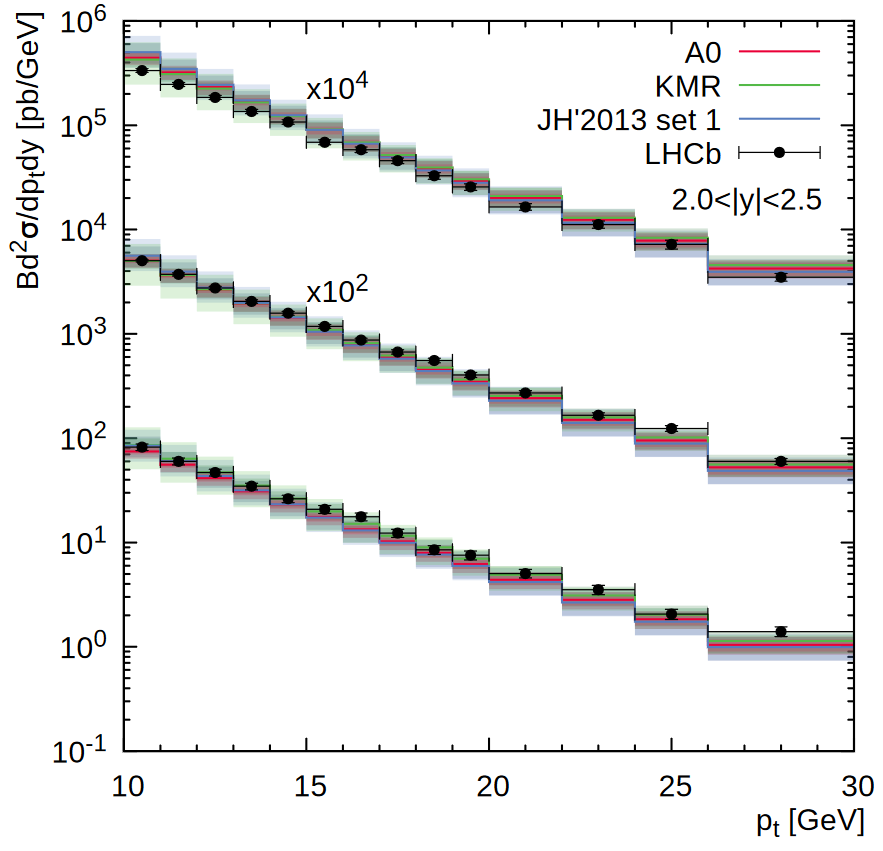}
\includegraphics[width=7.05cm]{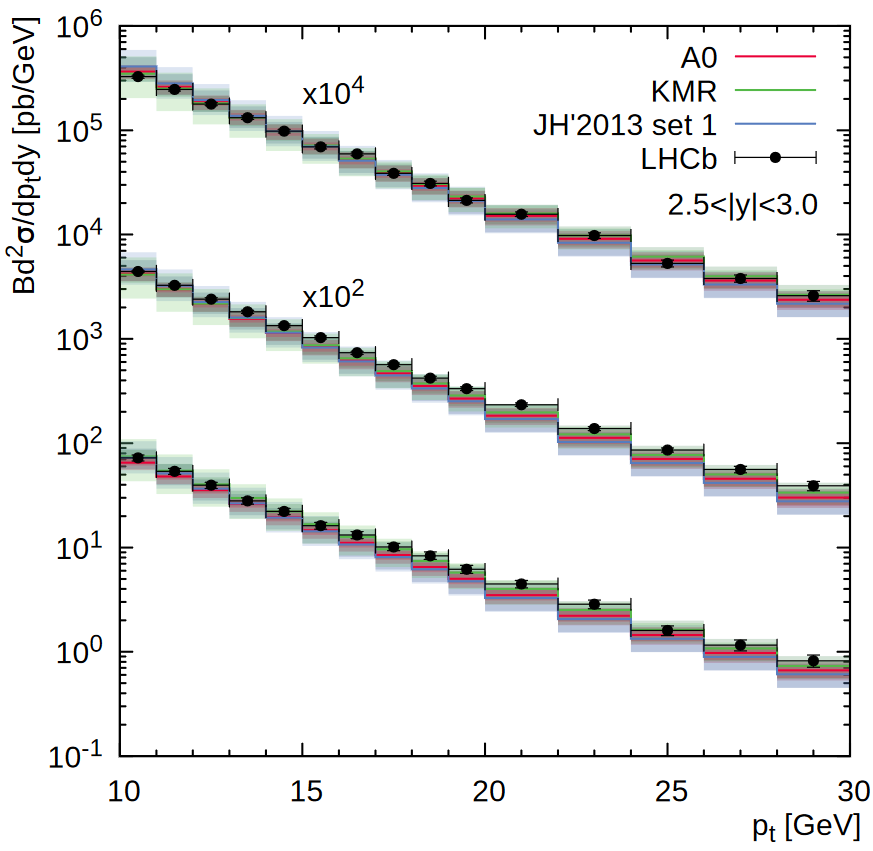}
\includegraphics[width=7.0cm]{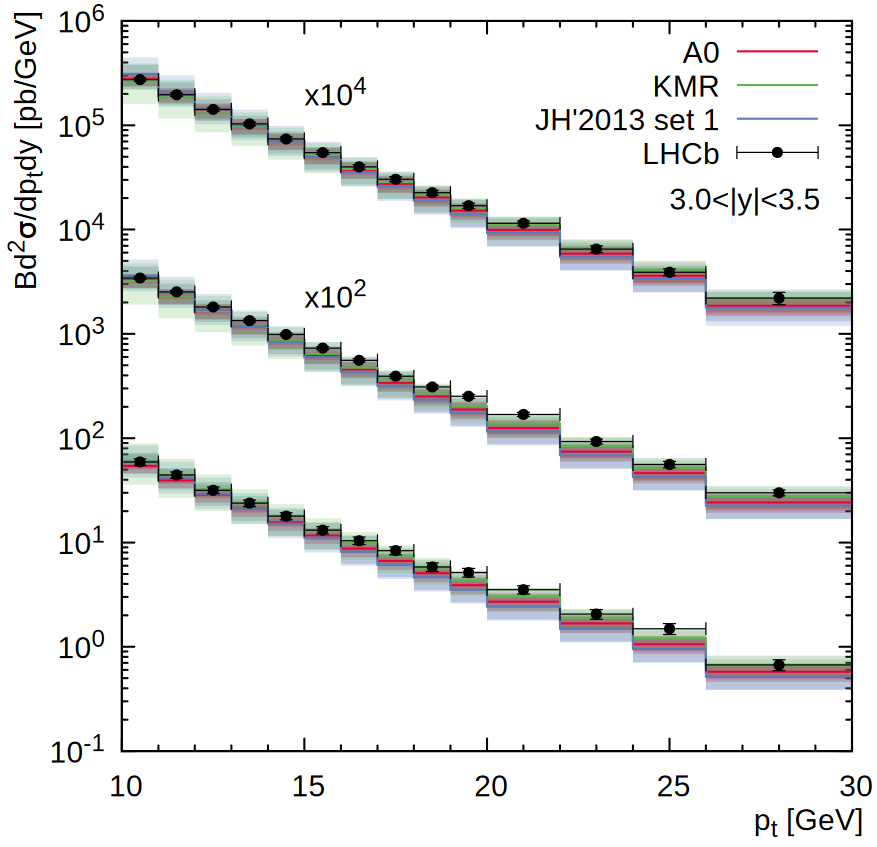}
\includegraphics[width=7.05cm]{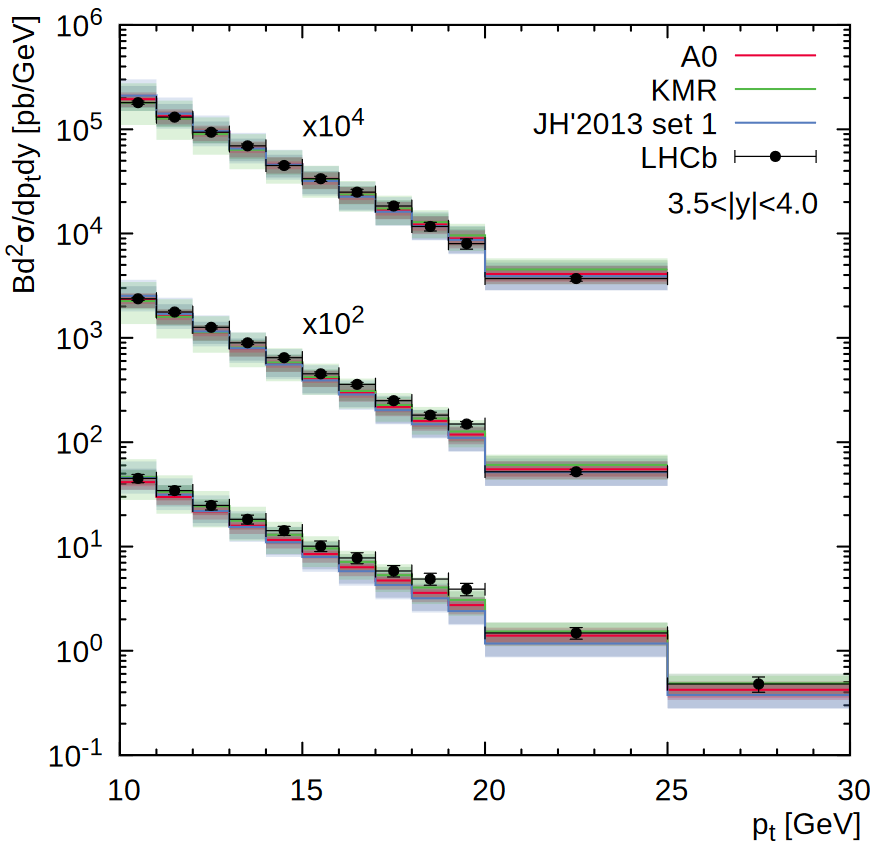}
\includegraphics[width=7.0cm]{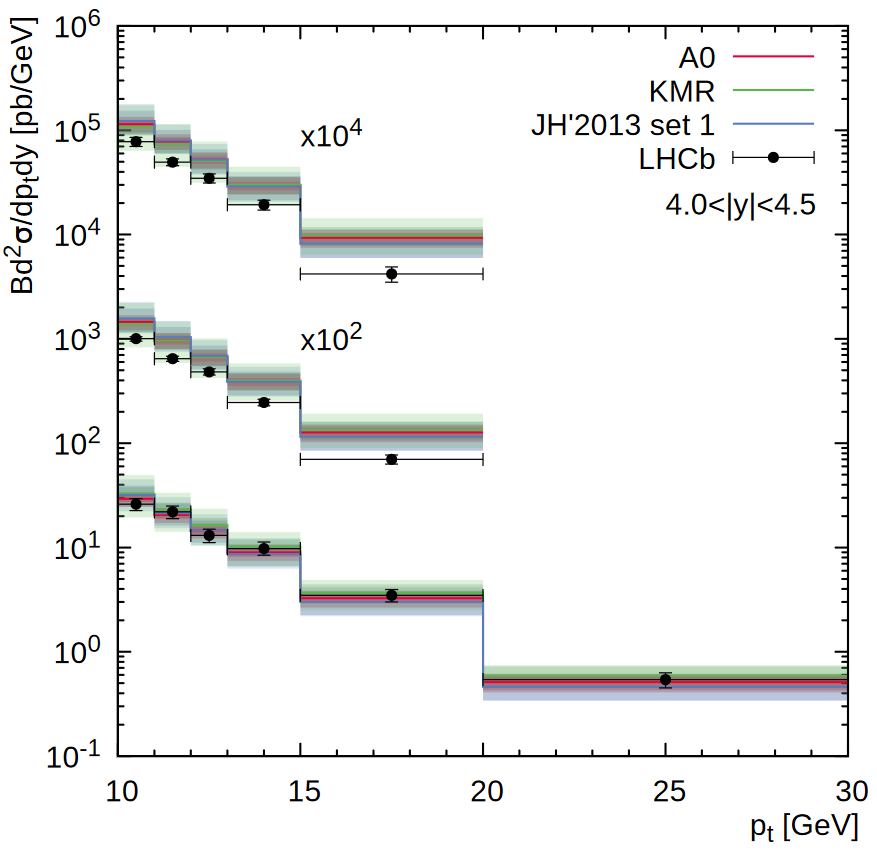}
\includegraphics[width=7.1cm]{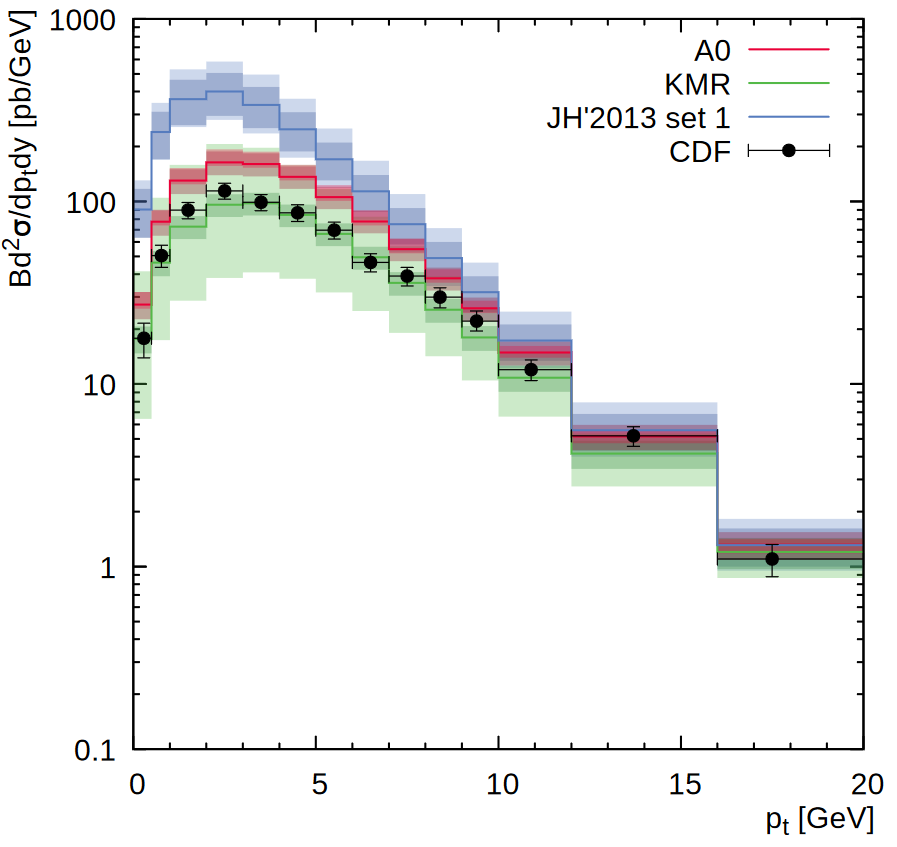}
\caption{Transverse momentum distribution of the  
  inclusive $\Upsilon(1S)$ production calculated at $\sqrt s = 1.8$, 
  $7$, $8$ and $13$~TeV in the different rapidity regions. 
  The notation of all histograms is the same as in Fig.~\ref{fig1}.
  The experimental data are from CDF \cite{cdf} and LHCb \cite{lhcb1,lhcb2}.}
\label{fig5}
\end{center}
\end{figure}

In addition, we have checked our results with the data, not included into the fit procedure: namely,
the rather old CDF data~\cite{cdf} taken at $\sqrt s = 1.8$~TeV
and the LHCb data~\cite{lhcb1,lhcb2} taken in the forward rapidity region $2 < y < 4.5$
at $\sqrt s = 7$, $8$ and $13$~TeV (see Fig.~\ref{fig5}). 
As one can see, we acceptably describe all the data above.
Moreover, we find that the KMR gluon density
does a much better job here than the JH'2013 set 1 or A0 distributions.
Remarkably, the KMR gluon is only one TMD gluon density which is able to 
reproduce well the measurements in the low $p_T$ region.

\begin{figure}
\begin{center}
\includegraphics[width=7.0cm]{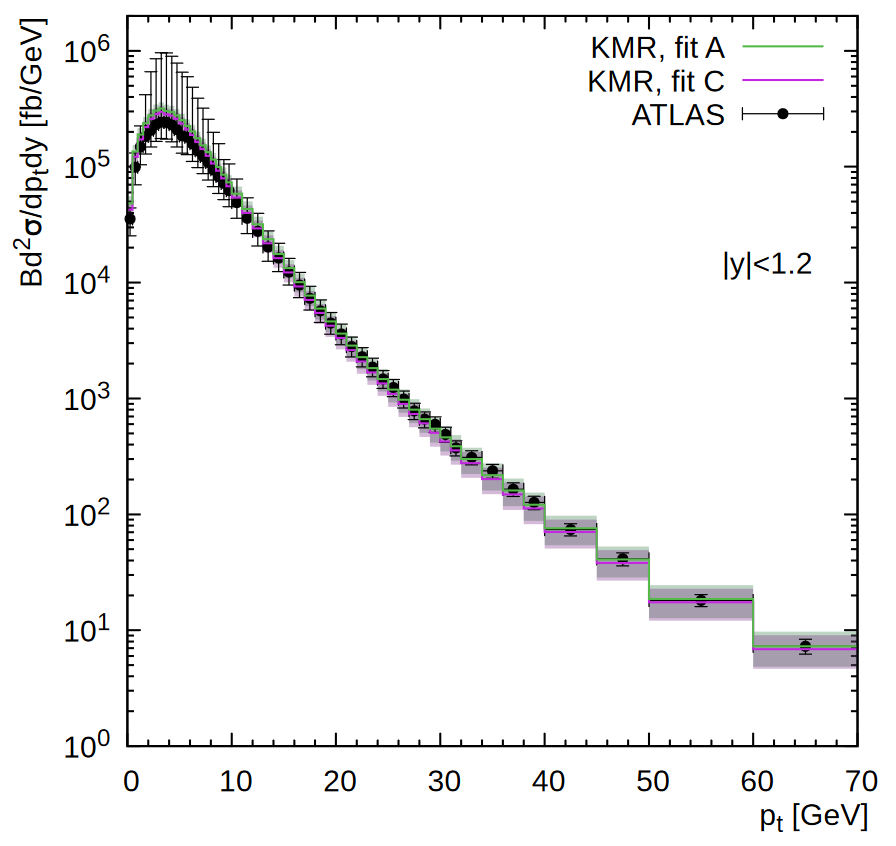}
\includegraphics[width=7.0cm]{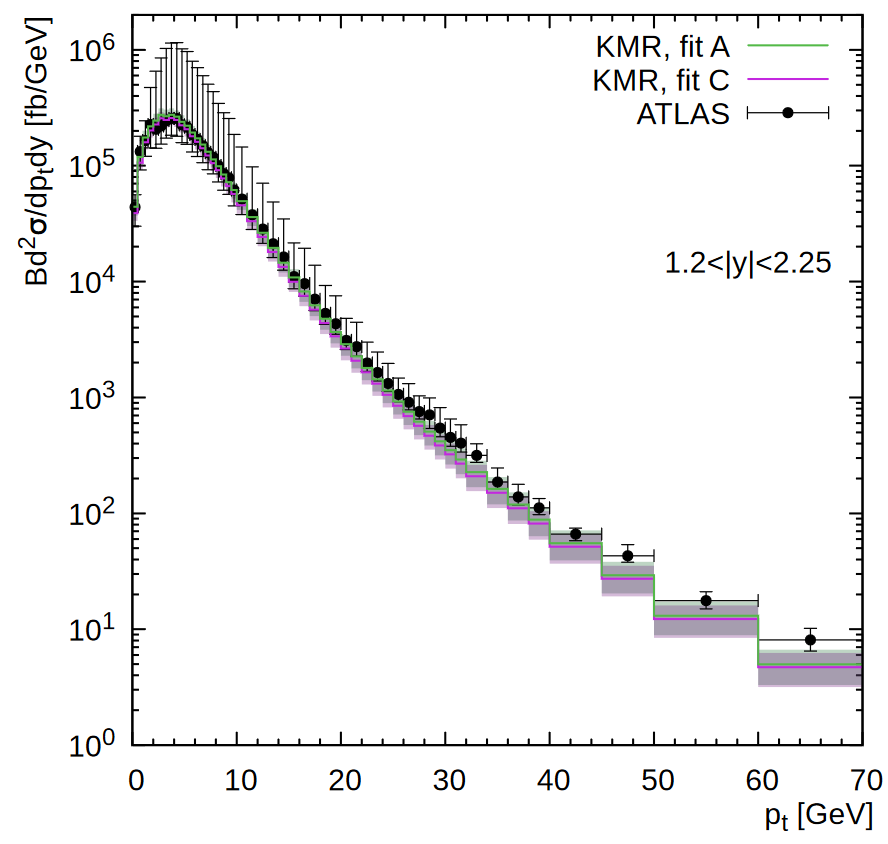}
\caption{Transverse momentum distribution of the 
  inclusive $\Upsilon(1S)$ production calculated at 
  $\sqrt s = 7$~TeV in the different rapidity regions. 
  The green and purple histograms
  correspond to the "fit A" and "fit C" predictions obtained with the KMR gluon density. 
  Shaded bands represent the NME uncertainties 
  of our calculations, as it is described in the text.
  The experimental data are from ATLAS\cite{atlas}.}
\label{fig10}
\end{center}
\end{figure}

\begin{figure}
\begin{center}
\includegraphics[width=7.0cm]{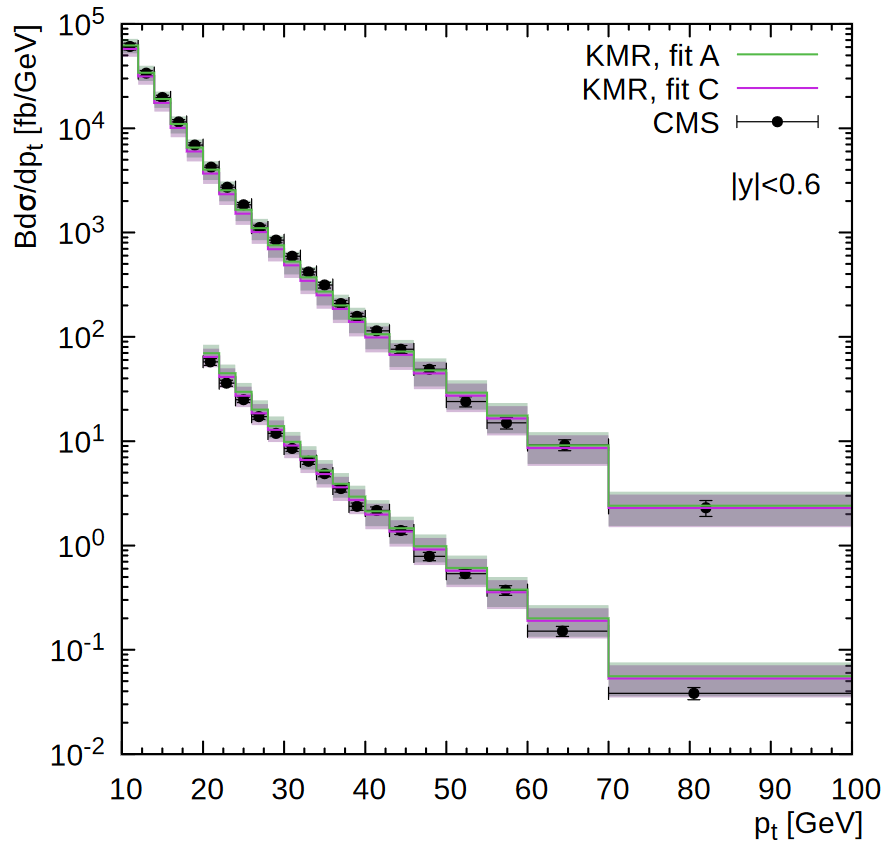}
\includegraphics[width=7.0cm]{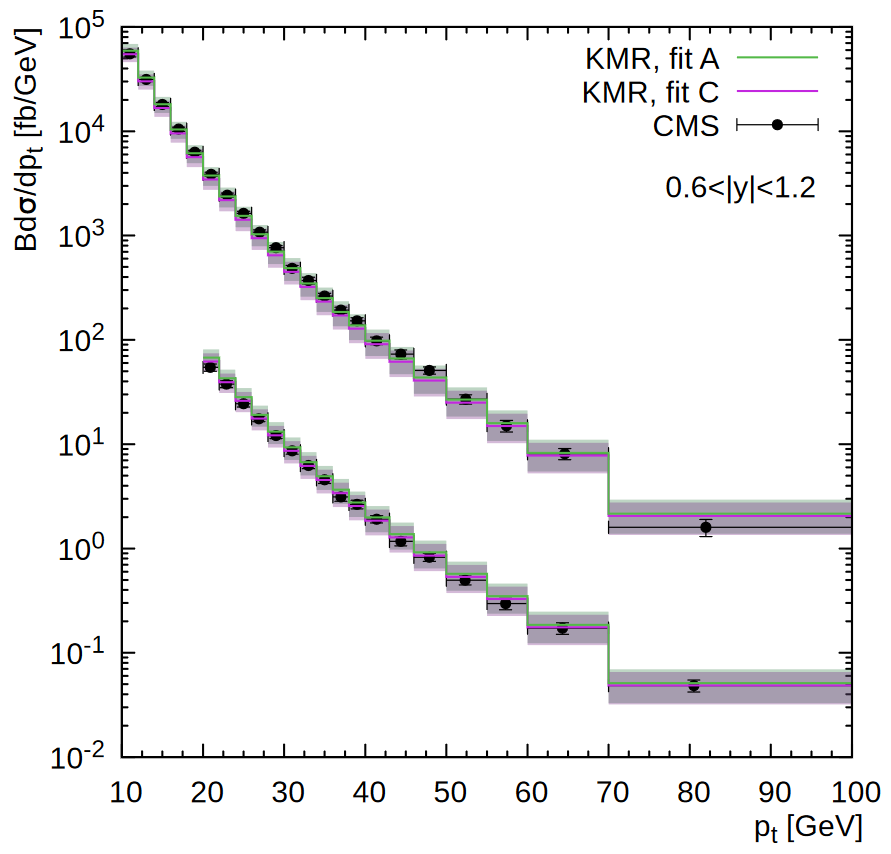}
\includegraphics[width=7.0cm]{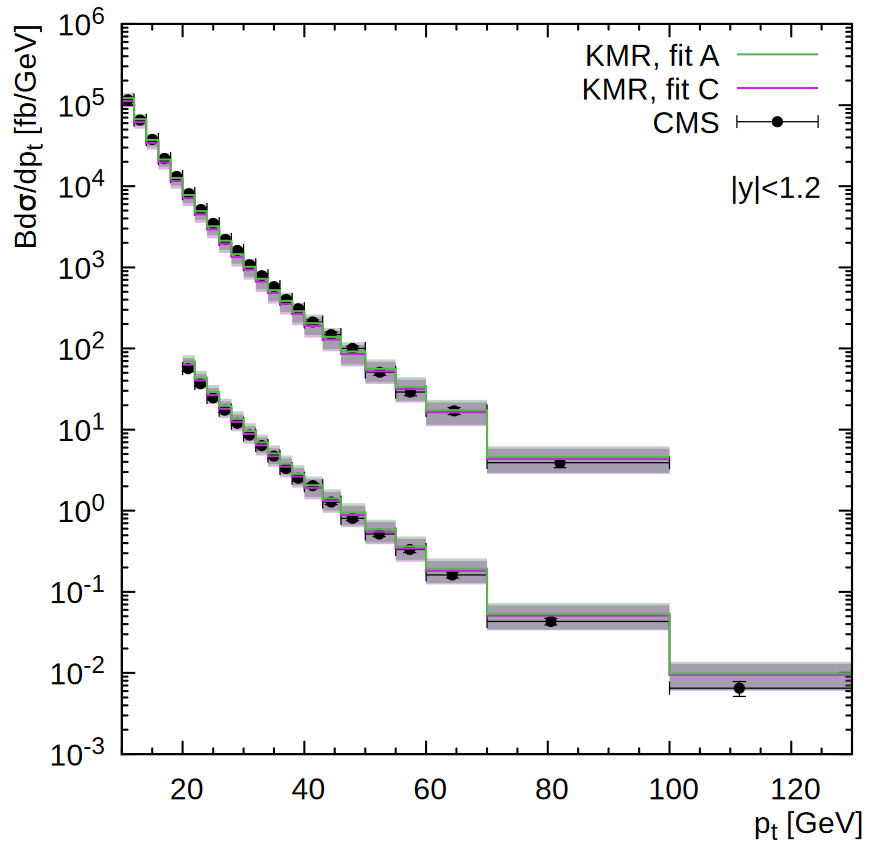}
\caption{Transverse momentum distribution of the
  inclusive $\Upsilon(1S)$ production calculated at $\sqrt s = 7$~TeV
  (upper histograms) and $\sqrt s = 13$~TeV (lower histograms, 
  divided by $100$) in the different rapidity regions. 
  The notation of all histograms is the same as in Fig.~\ref{fig10}.
  The experimental data are from CMS\cite{cms1,cms2}.}
\label{fig11}
\end{center}
\end{figure}

\begin{figure}
\begin{center}
\includegraphics[width=7.0cm]{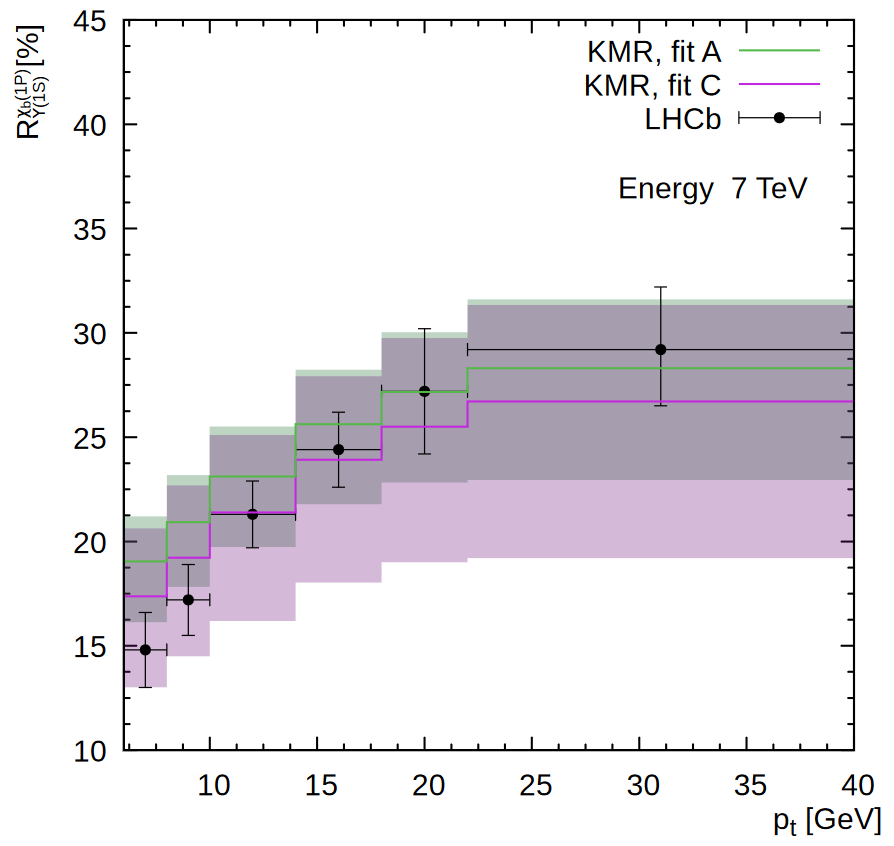}
\includegraphics[width=7.0cm]{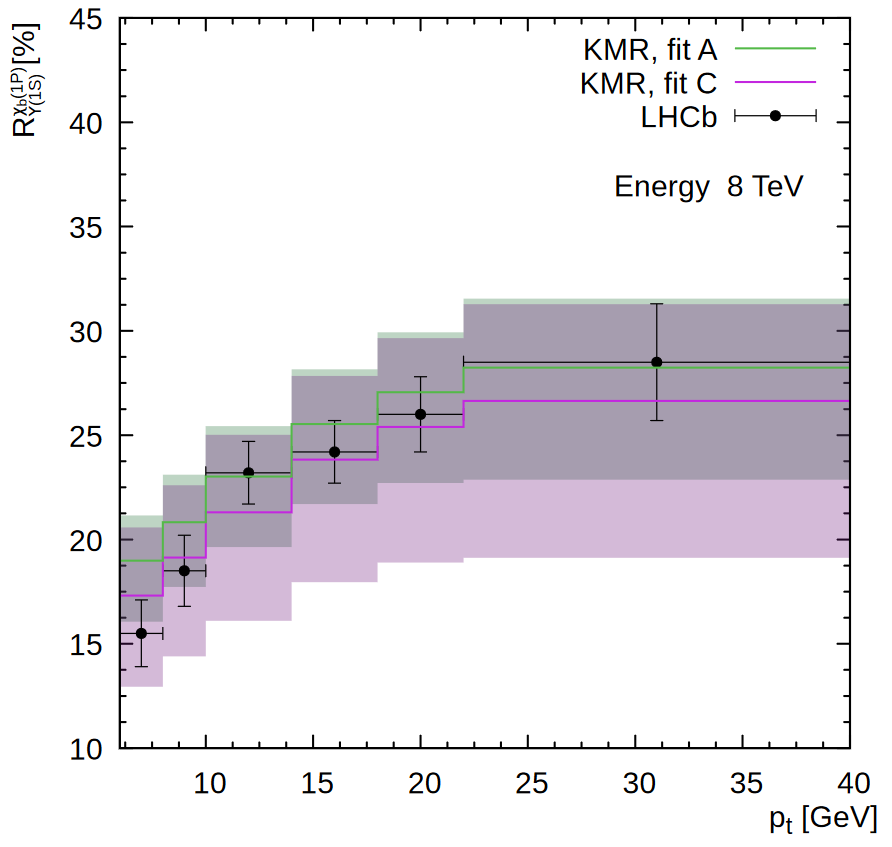}
\includegraphics[width=7.0cm]{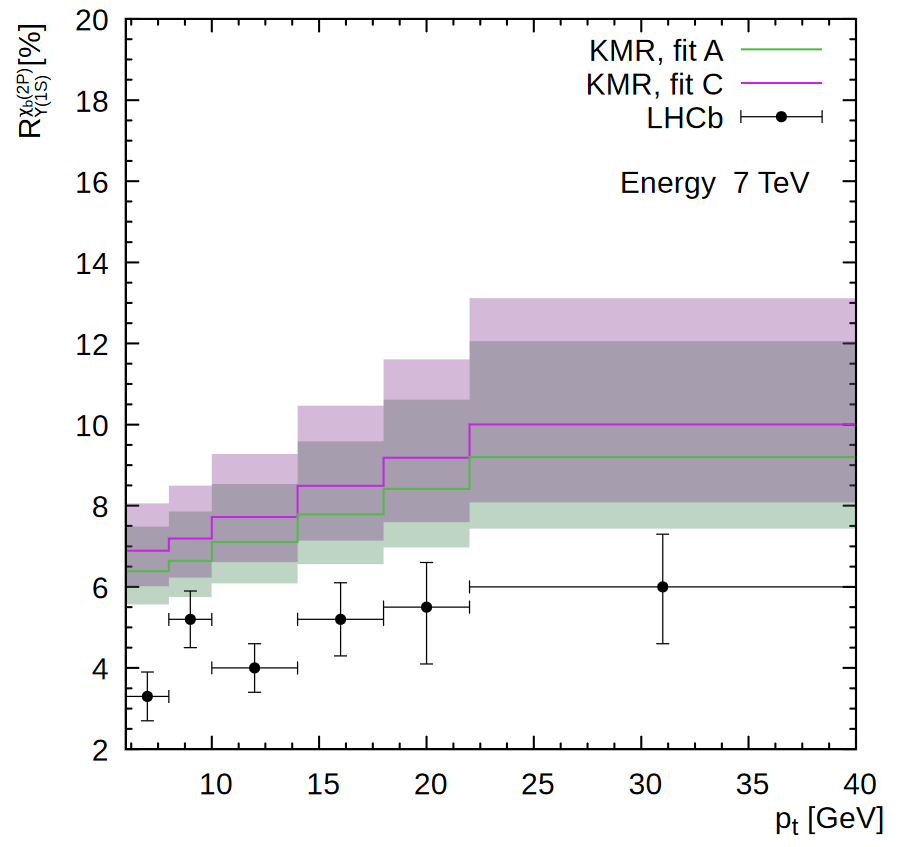}
\includegraphics[width=7.0cm]{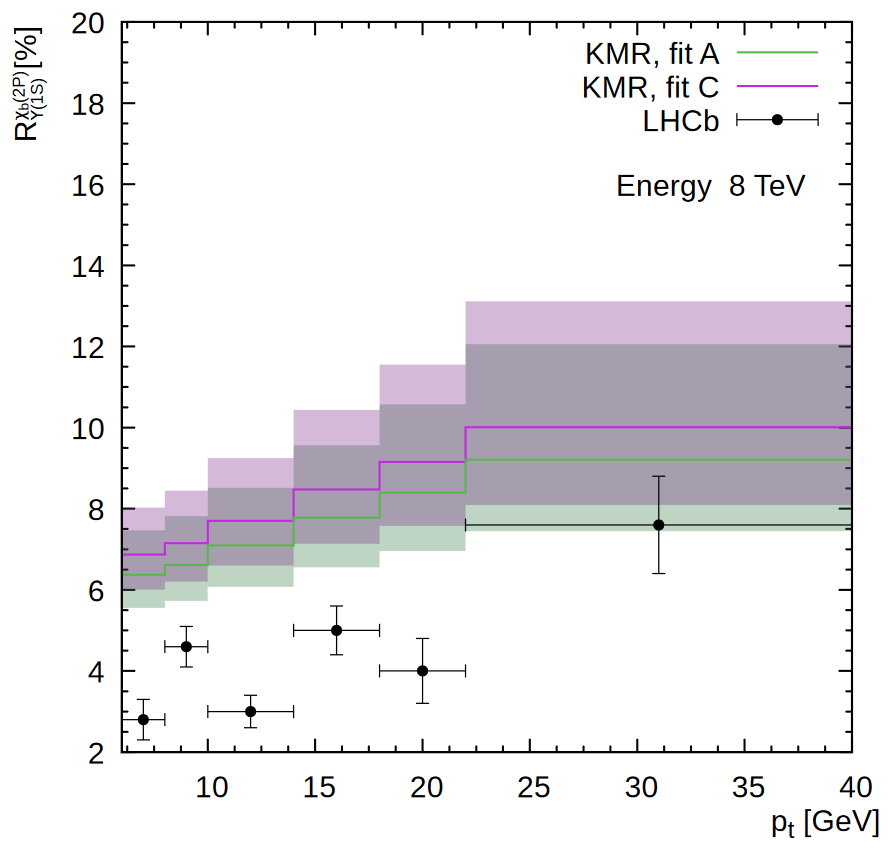}
\includegraphics[width=7.0cm]{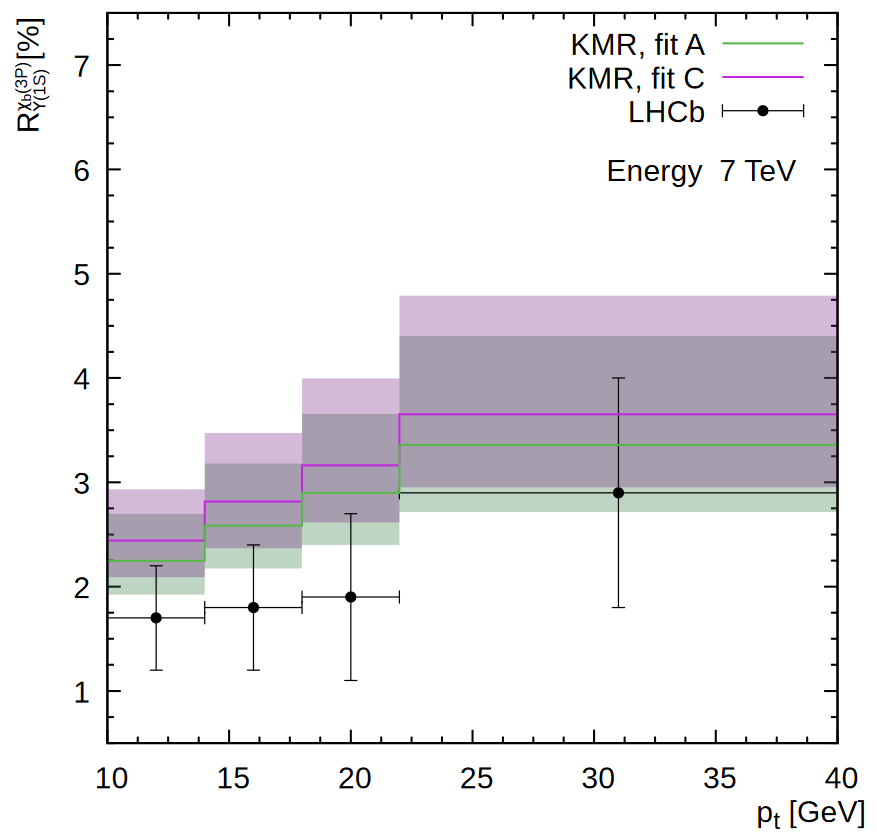}
\includegraphics[width=7.0cm]{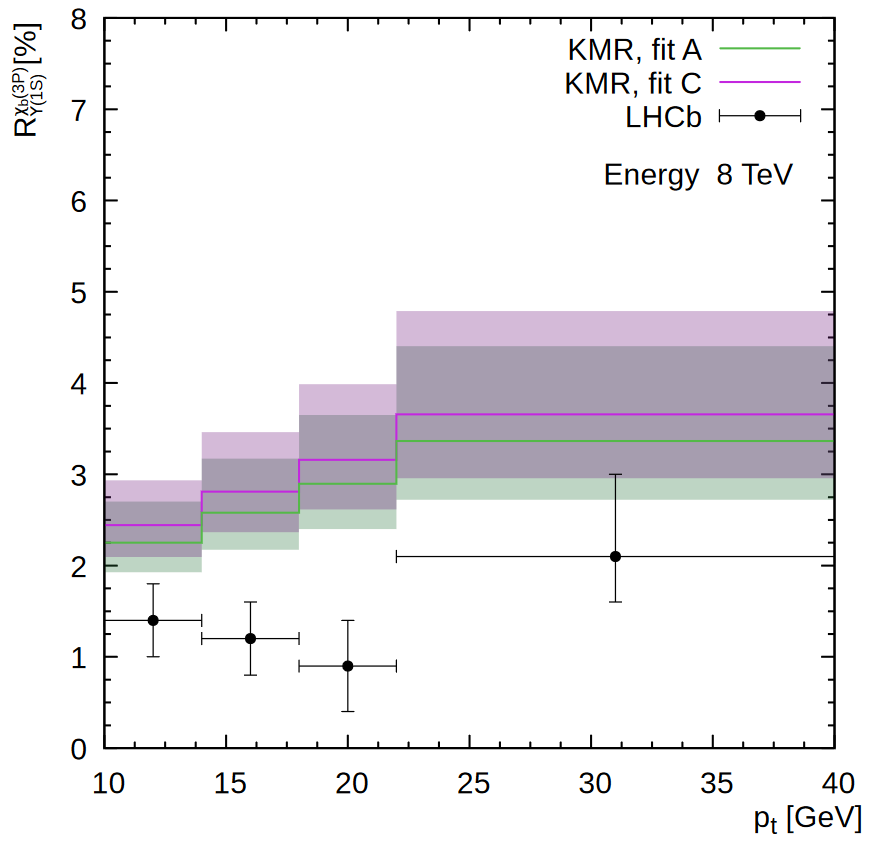}
\caption{The ratio $R^{\chi_b(mP)}_{\Upsilon(1S)}$
  calculated as a function of the $\Upsilon(1S)$ transverse momentum 
  calculated at $\sqrt s = 7$ and $8$~TeV.
  The notation of all histograms is the same as in Fig.~\ref{fig10}.
  The experimental data are from LHCb\cite{lhcbr}.}
\label{fig12}
\end{center}
\end{figure}

\begin{figure}
\begin{center}
\includegraphics[width=7.0cm]{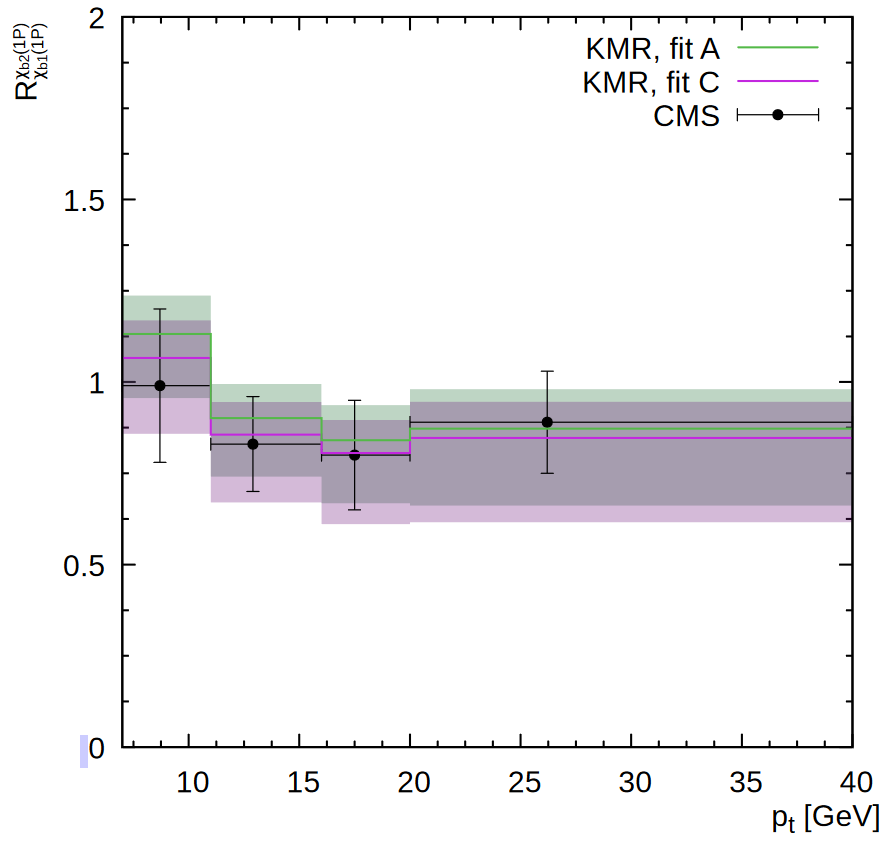}
\includegraphics[width=7.15cm]{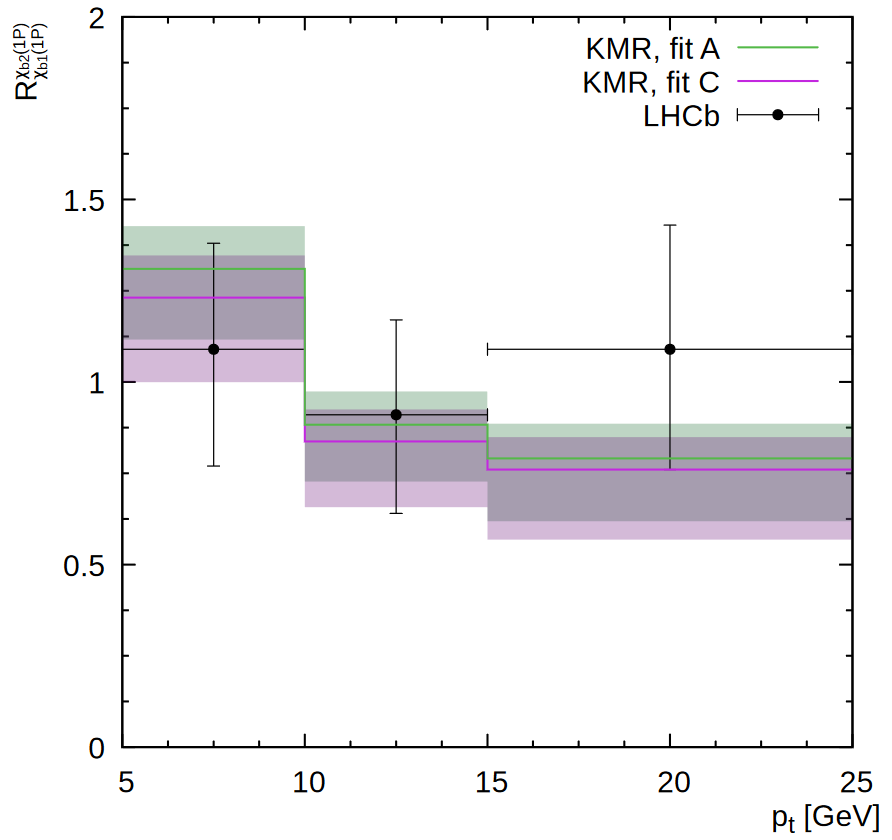}
\caption{The ratio $R^{\chi_{b2}(1P)}_{\chi_{b1}(1P)}$
  calculated as a function of the $\Upsilon(1S)$ transverse momentum 
  calculated at $\sqrt s = 8$~TeV.
  The notation of all histograms is the same as in Fig.~\ref{fig10}.
  The experimental data are from CMS\cite{cmsb2b1} and LHCb\cite{lhcbb2b1}.}
\label{fig13}
\end{center}
\end{figure}

\begin{figure}
\begin{center}
\includegraphics[width=7.0cm]{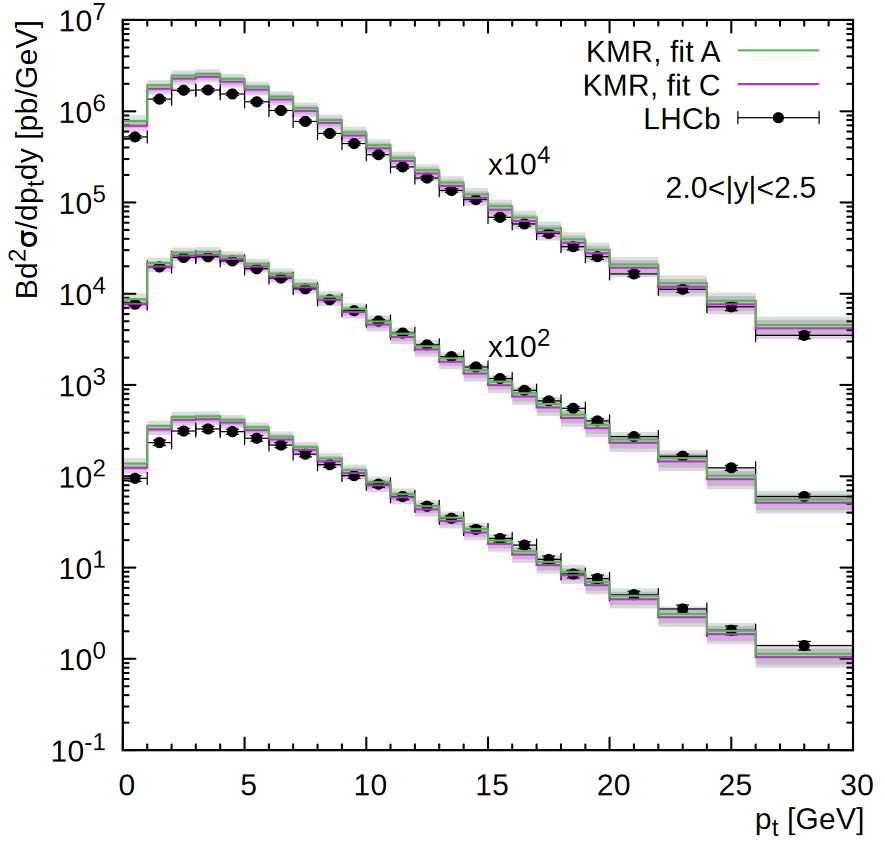}
\includegraphics[width=7.05cm]{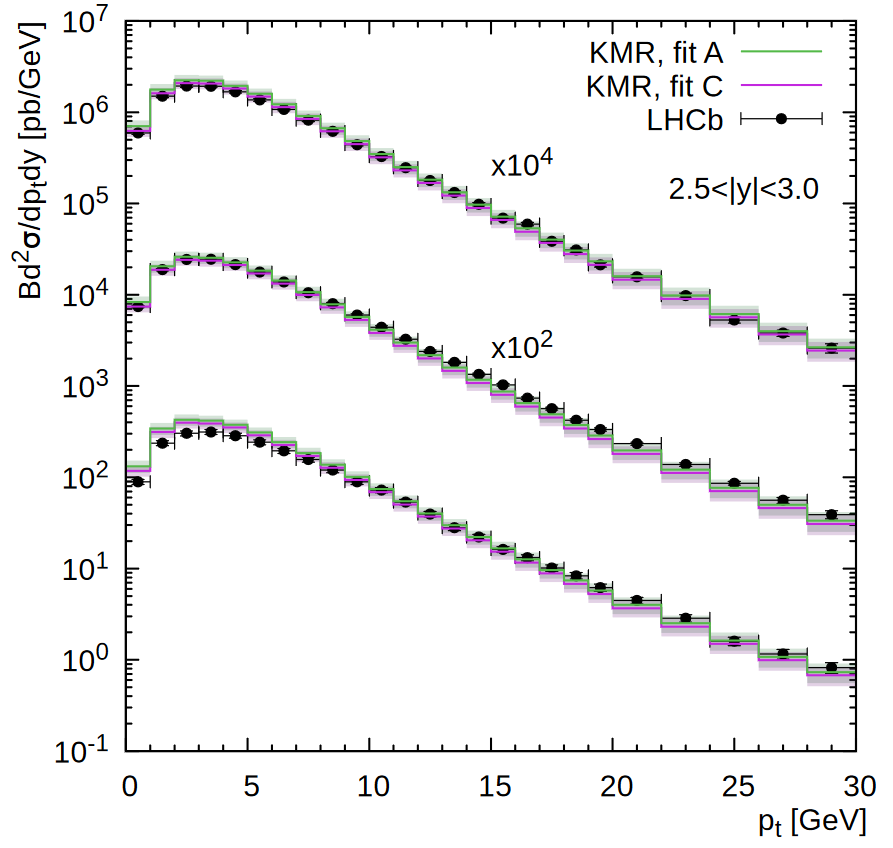}
\includegraphics[width=7.0cm]{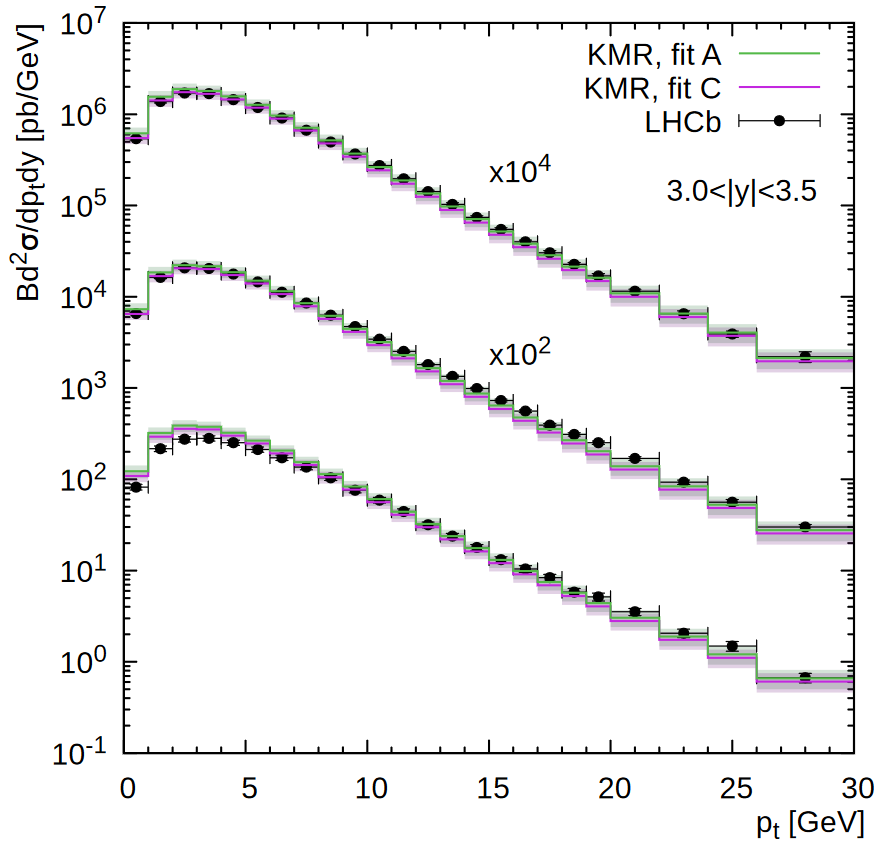}
\includegraphics[width=7.05cm]{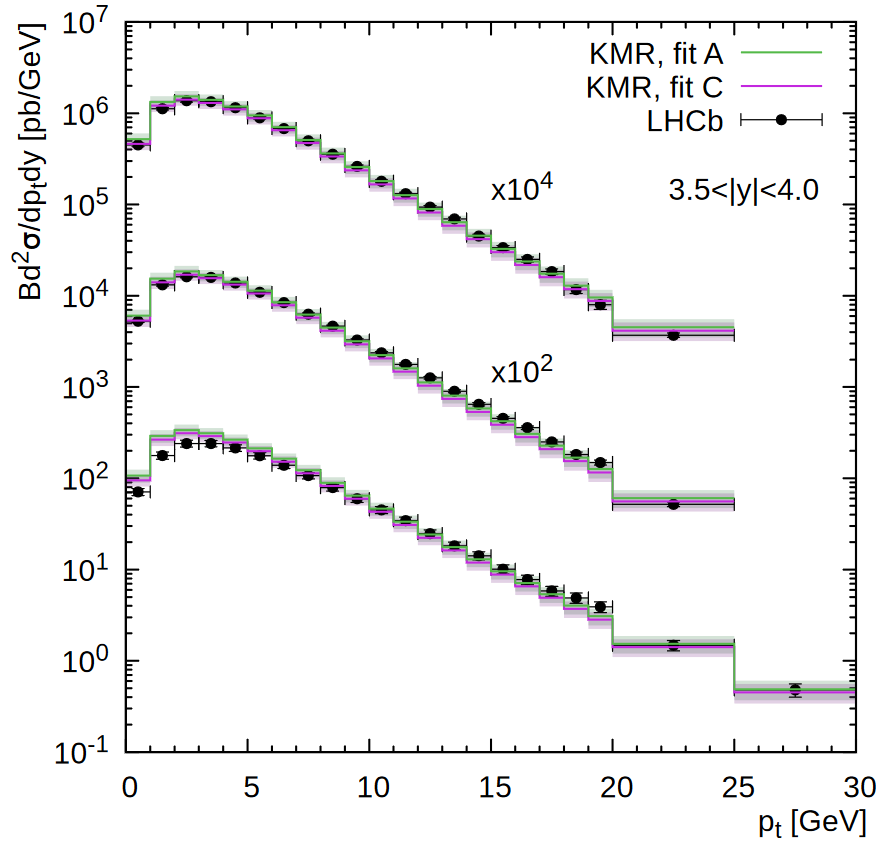}
\includegraphics[width=7.0cm]{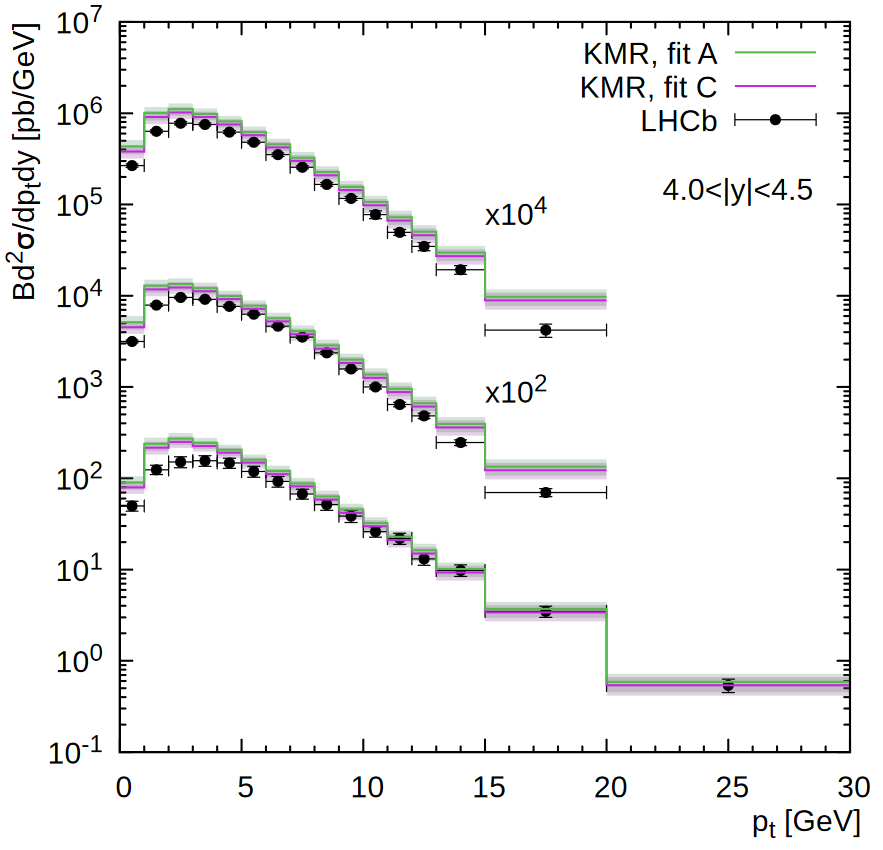}
\includegraphics[width=7.1cm]{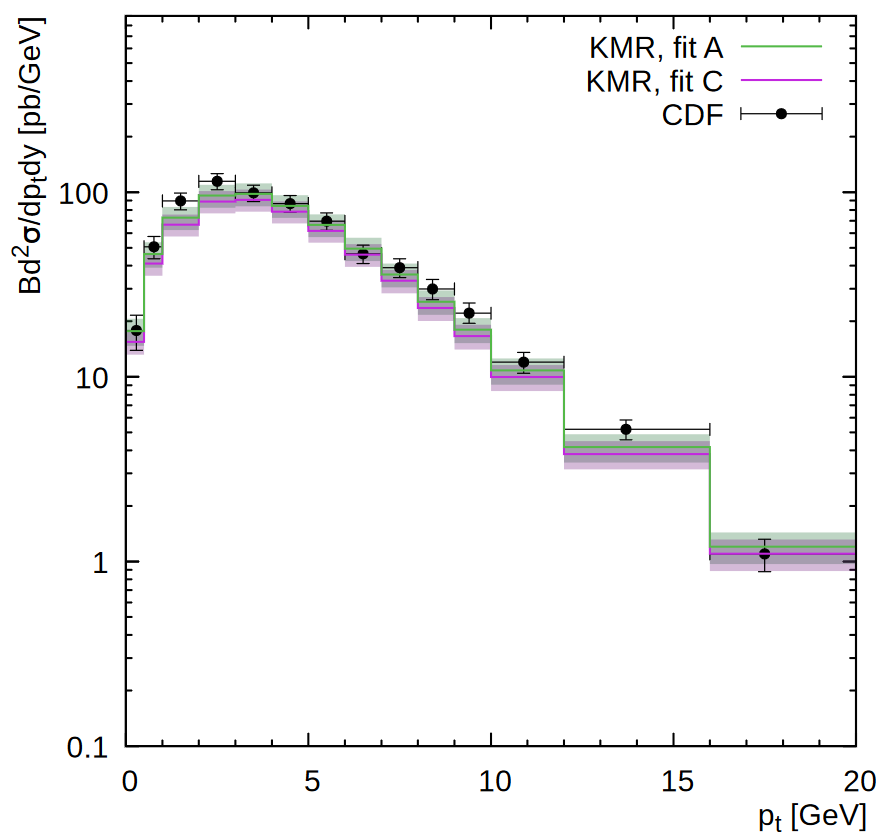}
\caption{Transverse momentum distribution of the  
  inclusive $\Upsilon(1S)$ production calculated at $\sqrt s = 1.8$, 
  $7$, $8$ and $13$~TeV in the different rapidity regions. 
  The notation of all histograms is the same as in Fig.~\ref{fig10}.
  The experimental data are from CDF\cite{cdf} and LHCb\cite{lhcb1,lhcb2}.}
\label{fig14}
\end{center}
\end{figure}

Based on that we have investigated the sensitivity of our fit to the low $p_T$ region using the KMR gluon density function. 
To do this we include the low $p_T$ region into our fit ("fit C"). 
Our results can be find in Table~\ref{tab1.5} and Figs.~\ref{fig10} --- \ref{fig14}, where we used 
the KMR gluon density using the NMEs from the "fit A'' and "fit C''. 
One can see that both fit scenarios
give overall almost the same results, and only for the
ratios $R^{\chi_b(2P)}_{\Upsilon(1S)}$ and $R^{\chi_b(3P)}_{\Upsilon(1S)}$
the "fit A'' gives slightly better results. However, the uncertainties for the "fit C''
are larger than for the "fit A'' scenario. 
Additionally, we present 
the corresponding $\chi^2/d.o.f.$ using the NMEs from the "fit A'' for A0, JH'2013 set 1, KMR distributions 
and from the "fit C'' for KMR only, that are listed in Table~\ref{tab3}. 
We should note that using the ATLAS\cite{atlas} and CMS\cite{cms1,cms2} data give 
us lower values of $\chi^2/d.o.f.$ than using only the CMS data. This is due to large uncertainties of 
the ATLAS data in the low $p_T$ region.
One can see that $\chi^2/d.o.f.$ for the "fit C'' are larger not only compared with the results using the NMEs 
from the "fit A'' scenario in the low $p_T$ region,
but also compared with NMEs from Table~\ref{tab1} in the $p_T>10$ GeV region.
These results justify our exclusion of the low $p_T$ region from our fit.

Now we turn to the polarization of $\Upsilon(1S)$ mesons at the LHC conditions.
It is well known that
the polarization of any vector meson can be described 
with three parameters $\lambda_\theta$, $\lambda_\phi$ and 
$\lambda_{\theta\phi}$, which determine the spin density matrix of a 
meson decaying into a lepton pair and can be measured experimentally. 
The double differential angular distribution of the decay leptons 
can be written as~\cite{siglam}:	
\begin{equation}
  {{d\sigma}\over{d\cos\theta^*d\phi^*}} \sim {{1}\over{3+\lambda_\theta}}(1 + \lambda_\theta\cos^2\theta^* + \lambda_\phi\sin^2\theta^*\cos2\phi^* + \lambda_{\theta\phi}\sin2\theta^*\cos\phi^*), 
  \label{eqlam}
\end{equation}

\noindent where $\theta^*$ and $\phi^*$ are the polar and azimuthal 
angles of the decay lepton measured in the meson rest frame.
The case of 
$(\lambda_\theta$, $\lambda_\phi$, $\lambda_{\theta \phi}) = (0,0,0)$
corresponds to an unpolarized state, while 
$(\lambda_\theta$, $\lambda_\phi$, $\lambda_{\theta \phi}) = (1,0,0)$
and $(\lambda_\theta$, $\lambda_\phi$, $\lambda_{\theta \phi}) = (-1,0,0)$
refer to fully transverse and fully longitudinal polarizations. 
The CMS Collaboration has measured all of these polarization parameters for the $\Upsilon(1S)$ mesons
as functions of their transverse momentum 
in three complementary frames: the Collins-Soper, helicity and 
perpendicular helicity ones at $\sqrt{s} = 7$~TeV~\cite{cmslam}. 
The CDF Collaboration has measured the polarization parameters  
in the helicity frame at $\sqrt{s} = 1.96$~TeV~\cite{cdf2}.
The frame-independent parameter
$\tilde \lambda = (\lambda_\theta + 3\lambda_\phi)/(1 - \lambda_\phi)$ 
has been additionally studied.
As it was done previously~\cite{upsilonI,upsilonII}, to estimate $\lambda_\theta$, $\lambda_\phi$, 
$\lambda_{\theta\phi}$ and $\tilde \lambda$ we generally follow the experimental procedure. 
We collect the simulated events in the kinematical region defined by
the experimental setup, generate the decay lepton angular
distributions according to the production and decay matrix elements 
and then apply a three-parametric fit based on~(\ref{eqlam}).

\begin{figure}
\begin{center}
\includegraphics[width=7.0cm]{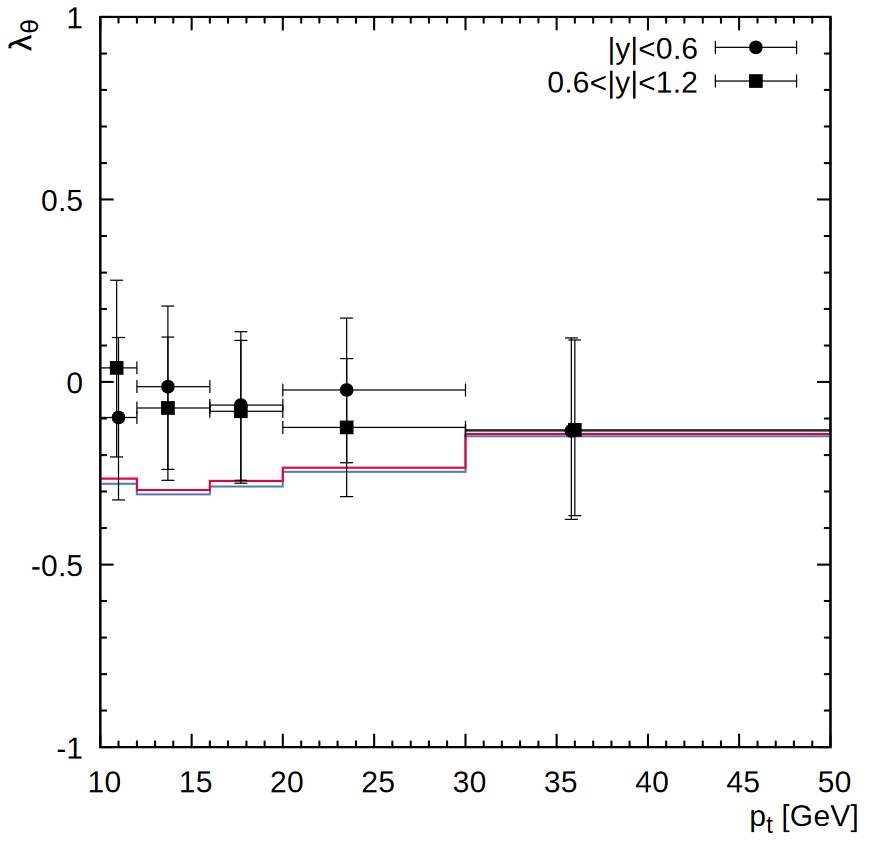}
\includegraphics[width=7.0cm]{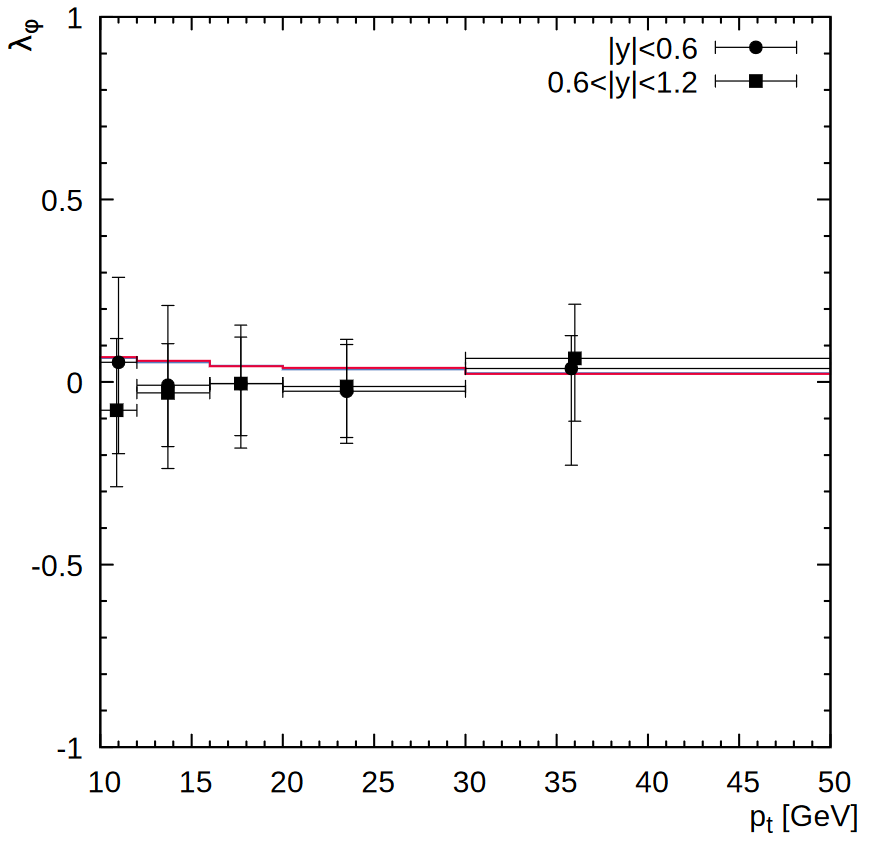}
\includegraphics[width=7.0cm]{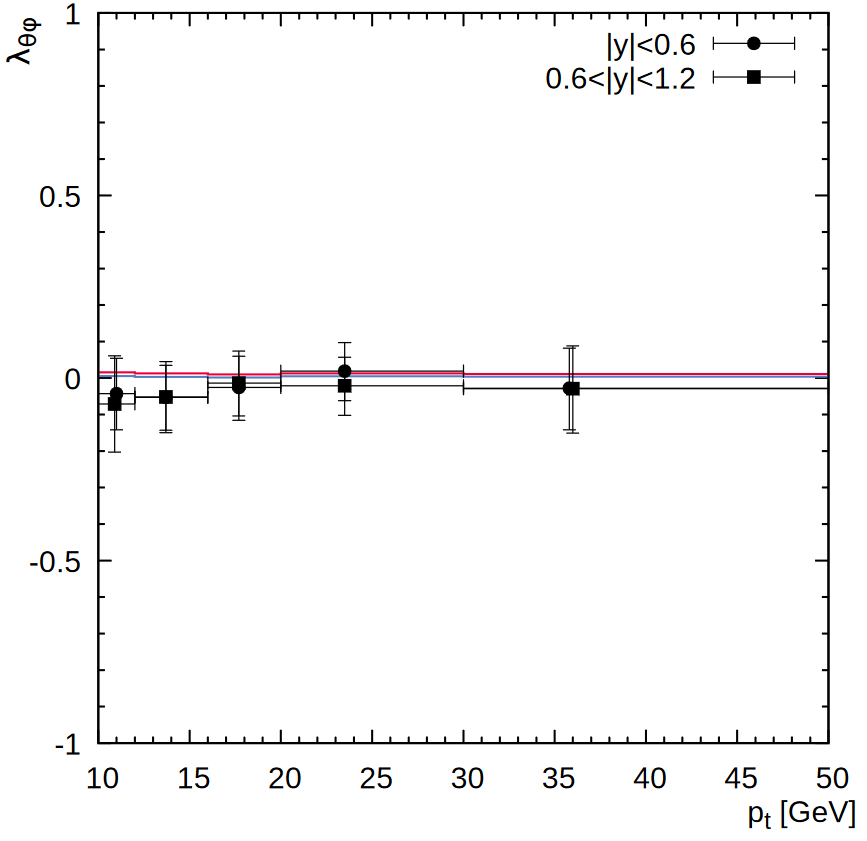}
\includegraphics[width=7.0cm]{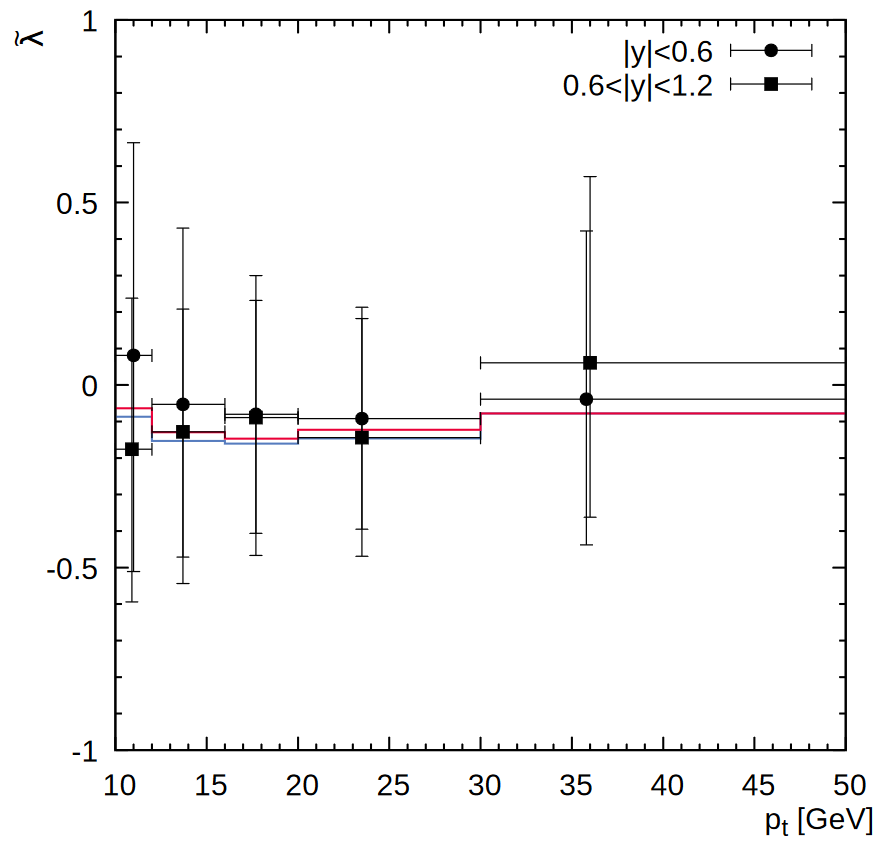}
\caption{The polarization parameters $\lambda_\theta$, 
  $\lambda_\phi$, $\lambda_{\theta\phi}$ and $\tilde\lambda$ of the 
  $\Upsilon(1S)$ mesons calculated in the CS frame as functions
  of its transverse momentum at $\sqrt{s} = 7$ TeV. 
  The A0 gluon density is used. The blue and red histograms 
  correspond to the predictions obtained at $|y|<0.6$ and 
  $0.6<|y|<1.2$, respectively. 
  The experimental data are from CMS\cite{cmslam}.}
\label{fig6}
\end{center}
\end{figure}

\begin{figure}
\begin{center}
\includegraphics[width=7.0cm]{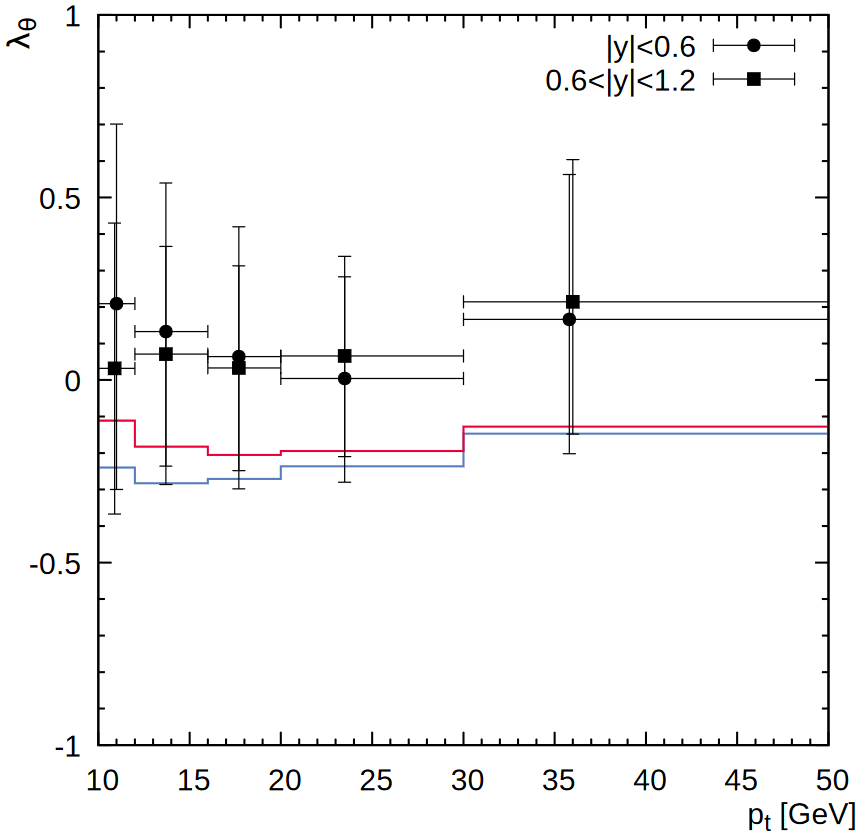}
\includegraphics[width=7.0cm]{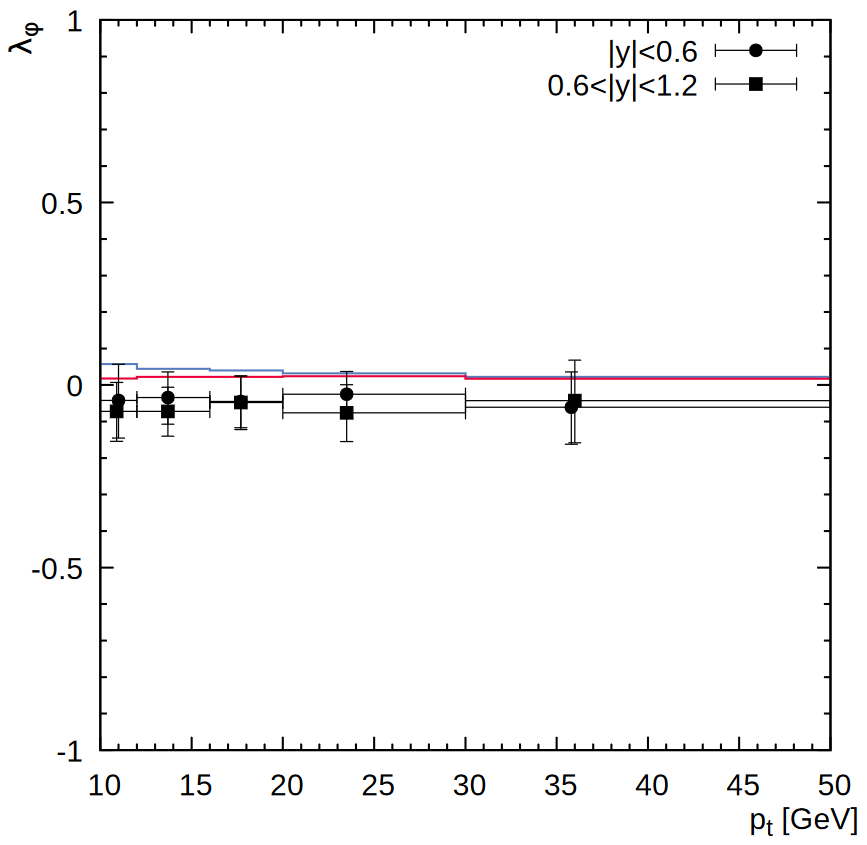}
\includegraphics[width=7.0cm]{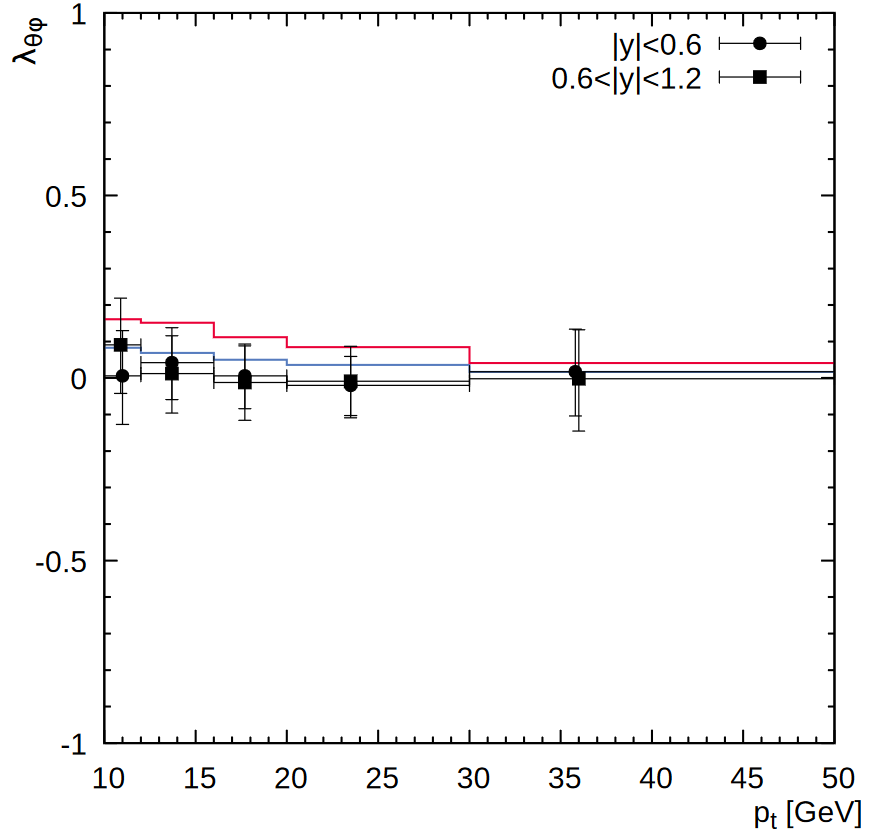}
\includegraphics[width=7.0cm]{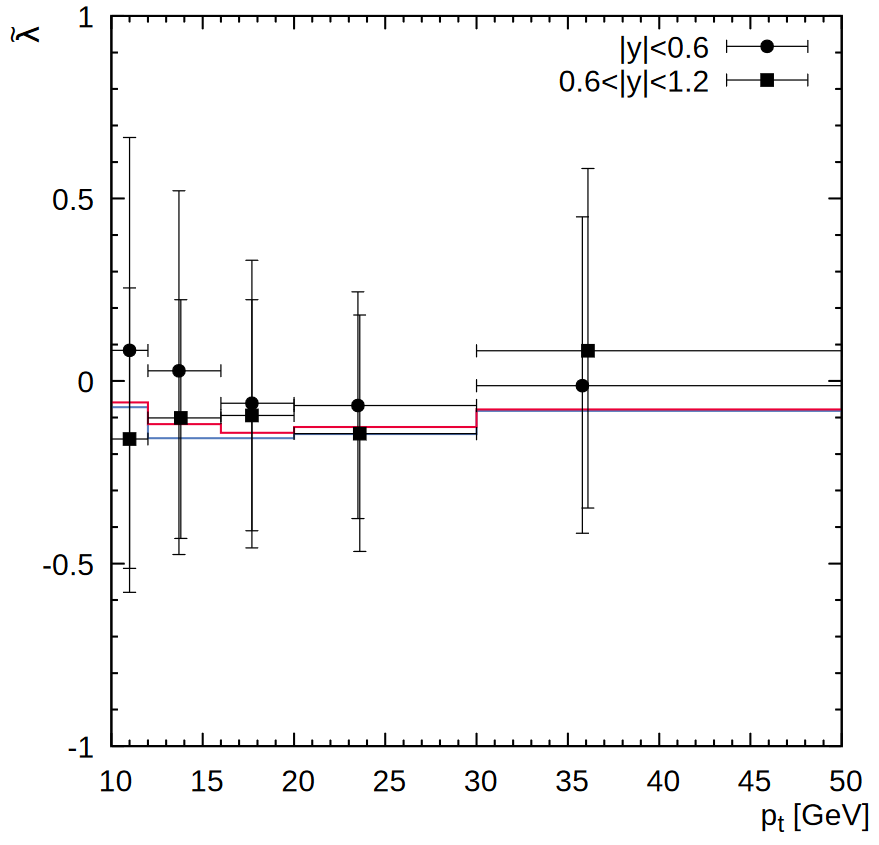}
\caption{The polarization parameters $\lambda_\theta$, 
  $\lambda_\phi$, $\lambda_{\theta\phi}$ and $\tilde\lambda$ of the
  $\Upsilon(1S)$ mesons calculated in the helicity frame as functions
  of its transverse momentum at $\sqrt{s} = 7$ TeV. 
  The notation of all histograms is the same as in Fig.~\ref{fig6}.
  The experimental data are from CMS\cite{cmslam}.}
\label{fig7}
\end{center}
\end{figure}

\begin{figure}
\begin{center}
\includegraphics[width=7.0cm]{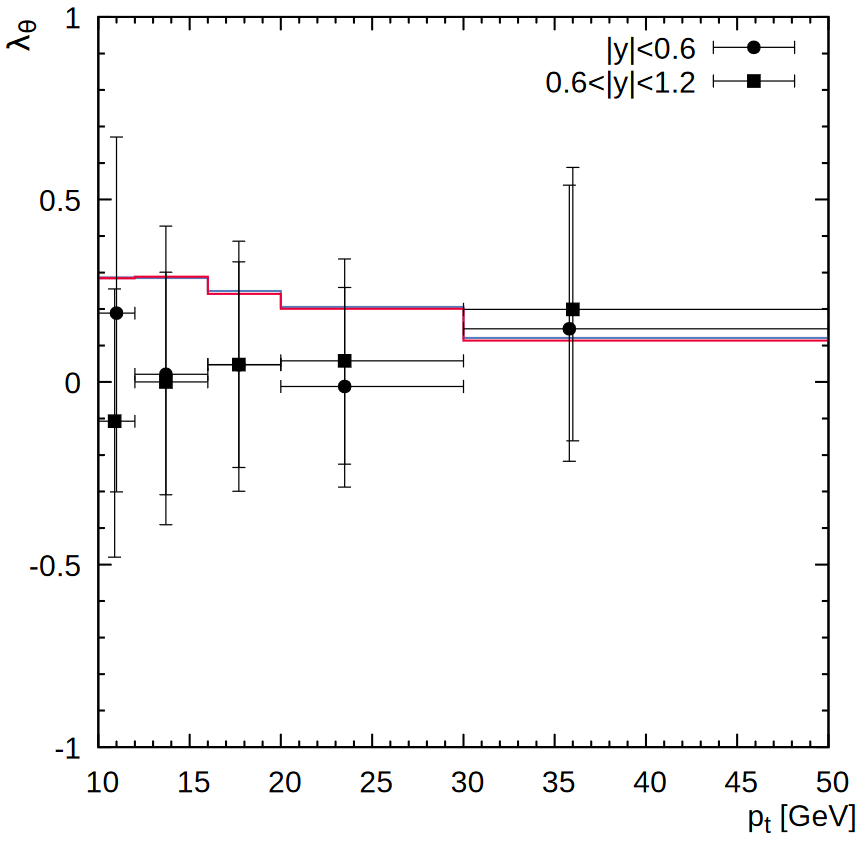}
\includegraphics[width=7.0cm]{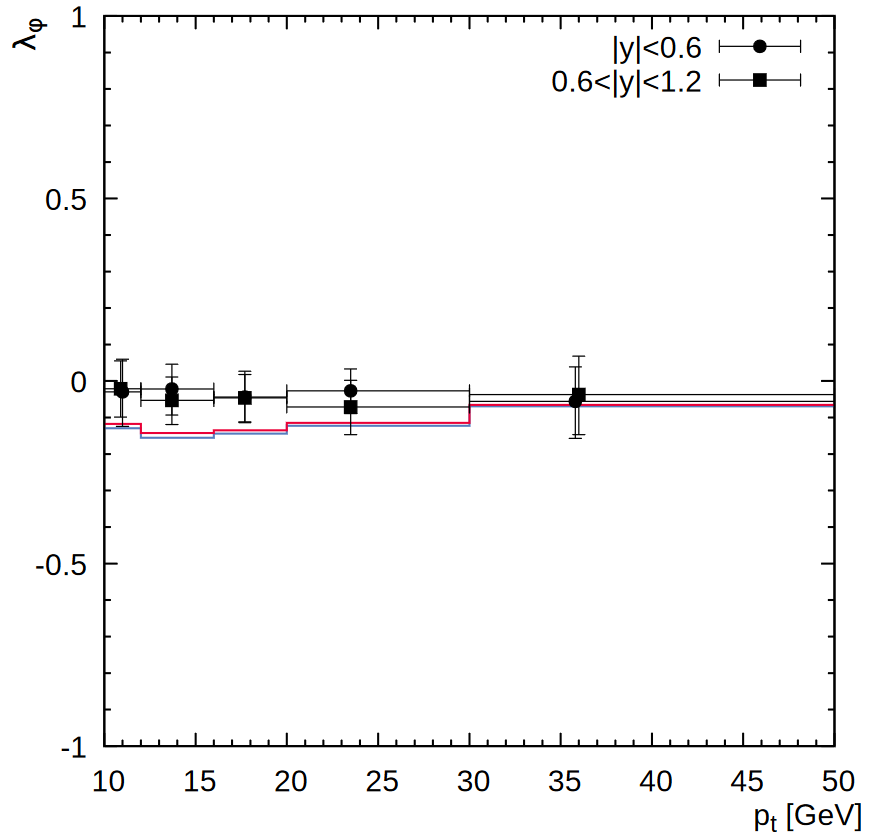}
\includegraphics[width=7.0cm]{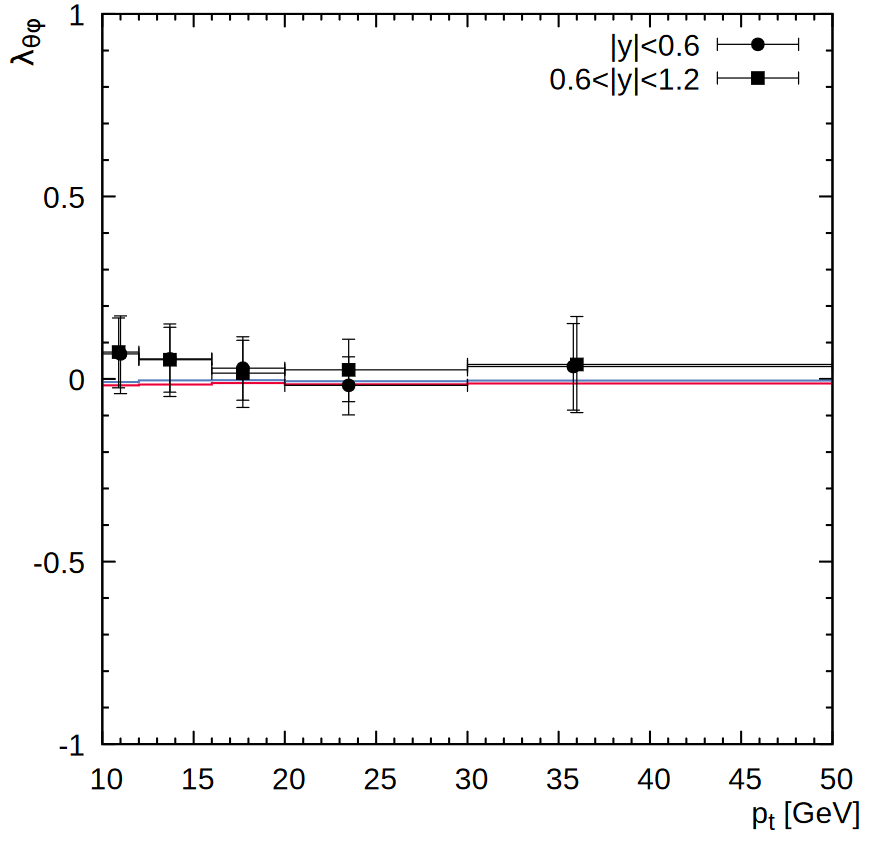}
\includegraphics[width=7.0cm]{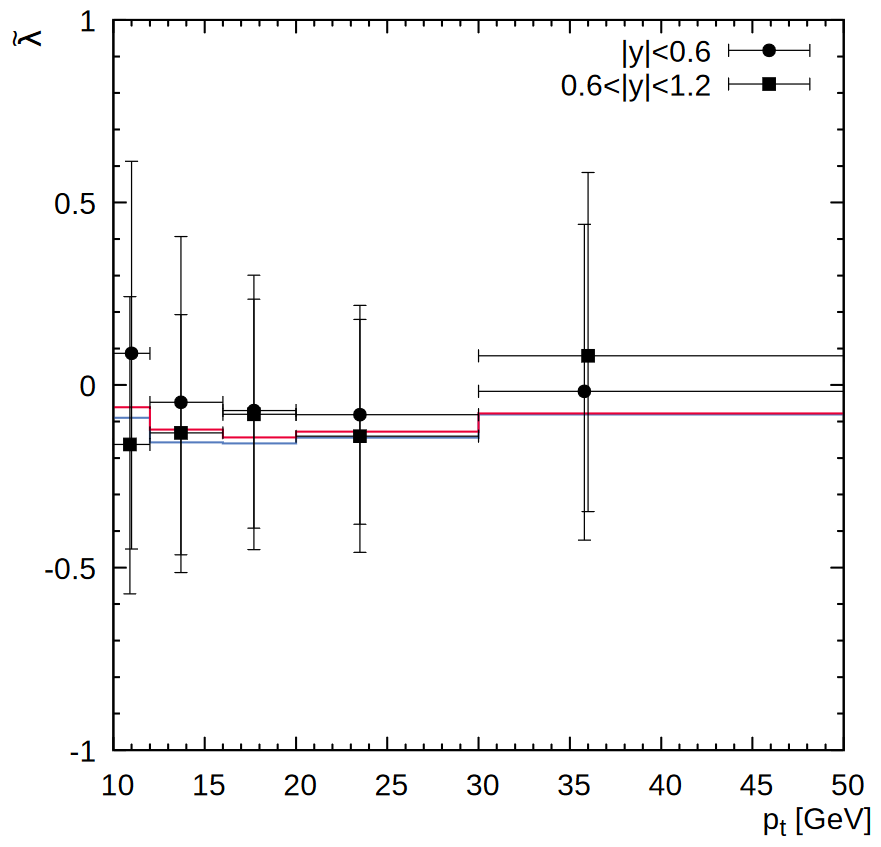}
\caption{The polarization parameters $\lambda_\theta$, 
  $\lambda_\phi$, $\lambda_{\theta\phi}$ and $\tilde\lambda$ of the 
  $\Upsilon(1S)$ mesons calculated in the perpendicular 
  helicity frame as functions
  of its transverse momentum at $\sqrt{s} = 7$ TeV. 
  The notation of all histograms is the same as in Fig.~\ref{fig6}.
  The experimental data are from CMS \cite{cmslam}.}
\label{fig8}
\end{center}
\end{figure}

\begin{figure}
\begin{center}
\includegraphics[width=7.0cm]{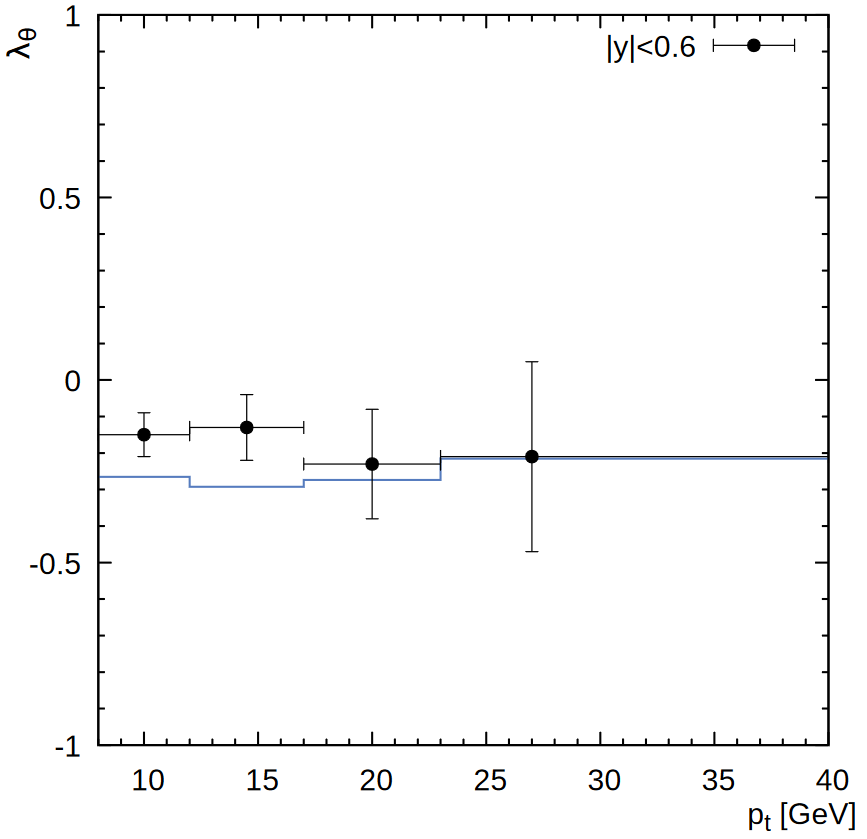}
\includegraphics[width=7.0cm]{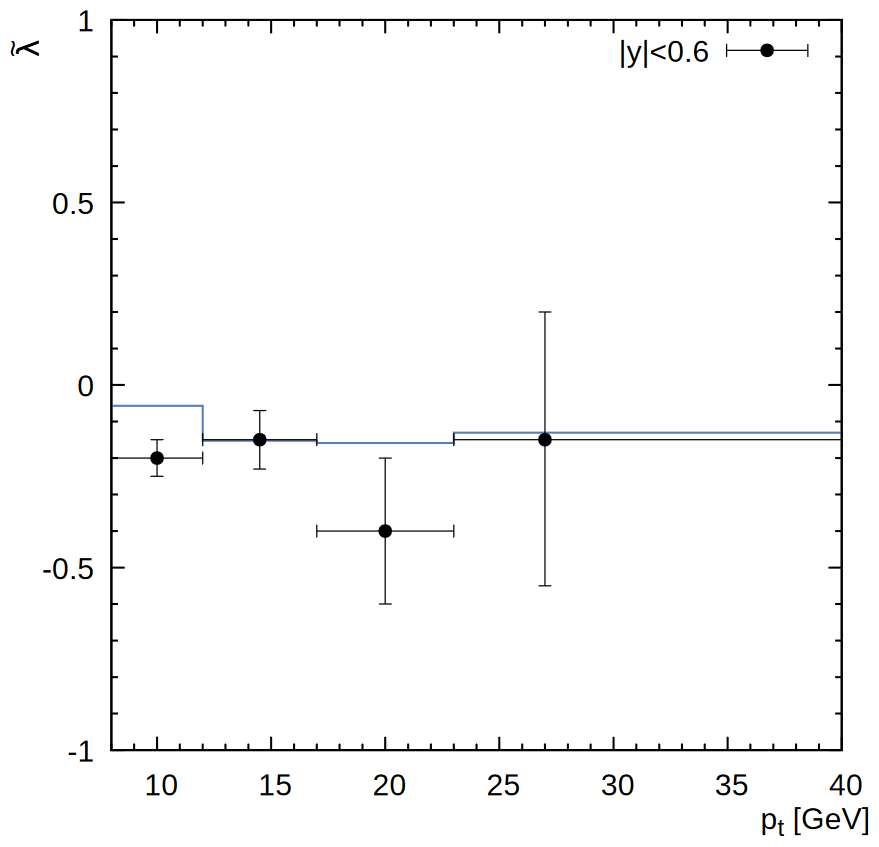}
\caption{The polarization parameters $\lambda_\theta$ and 
$\tilde\lambda$ of the $\Upsilon(1S)$ mesons calculated in the 
helicity frame as functions of its transverse momentum at 
$\sqrt{s} = 1.96$ TeV. The notation of all histograms is the same as in Fig.~\ref{fig6}.
  The experimental data are from CDF \cite{cdf2}.}
\label{fig9}
\end{center}
\end{figure}

Our predictions are shown in Figs.~\ref{fig6} --- \ref{fig9}. 
The calculations were performed using the A0
gluon density which provides the best description of the measured 
$\Upsilon(1S)$ transverse momenta distributions at the LHC conditions.
The NMEs from Table\ref{tab1} (the "fit A" scenario) were applied.
As one can see, we find only a weak or zero polarization in the all kinematic regions,
that perfectly agrees with the CMS and CDF measurements. 
This agreement shows no fundamental problems in describing the $\Upsilon(1S)$ 
polarization data.
Moreover, the calculated polarization parameters 
$\lambda_\theta$, $\lambda_\phi$, $\lambda_{\theta\phi}$ and $\tilde\lambda$
are stable with respect to variations in the model parameters. In fact,
there is no dependence on the strong coupling constant and/or TMD gluon densities in a proton.
As it was already pointed out above, our results 
for $\lambda_\theta$, $\lambda_\phi$, $\lambda_{\theta\phi}$ and $\tilde\lambda$
are based on the 
key assumption~\cite{baranov}
that the intermediate color octet states are states
with a definite total angular momentum $J$ and its projection $J_z$, rather than
states with definite projections of a spin $S_z$ and orbital angular momentum
$L_z$. Given that, the transition amplitudes only involve the polarization vector
associated with $J_z$ and not with $L_z$. As a result, we have no conservation
of $S_z$ in the electric dipole transitions. 
Under this assumption, 
we have achieved a reasonable simultaneous description 
for all of the available data for the $\Upsilon(1S)$ and $\chi_b(1P)$ mesons 
(the transverse momentum distributions, relative production rates
and polarization observables).
We have obtained earlier similar results for charmonia
($J/\psi$, $\psi^\prime$, $\chi_c$), 
$\Upsilon(2S)$ and $\Upsilon(3S)$ polarizations~\cite{baranovLip,baranovLip2,chicJrecent,upsilonI,upsilonII}.
Thus, keeping in mind the remarkable absence of tension with the $\eta_c$
production data (see\cite{baranovLip,baranovLip2}), one can conclude that the approach~\cite{baranov}
results in the self-consistent and simultaneous 
description of 
charmonium and bottomonium data
and therefore can be considered as providing an easy and natural solution to the long-standing
quarkonia production and polarization puzzle.

{\sl Acknowledgements}.
The authors thank S.P.~Baranov and M.A.~Malyshev for their 
interest, useful discussions and important remarks.
N.A.A. is supported by the Foundation for the Advancement of 
Theoretical Physics and Mathematics ``Basis'' (grant No.18-1-5-33-1) and by RFBR, project number 19-32-90096.
A.V.L. is grateful the DESY Directorate for the support
in the framework of Cooperation Agreement between MSU and DESY 
on phenomenology of the LHC processes and TMD parton densities.

{\bibliography{biblio}}
\newpage

	\begin{table}[H] \footnotesize
	\centering
	\begin{tabular}{lccccc}
	\hline
	\hline
	\\
	 &  A0 & JH'2013 set 1 &  KMR  & NLO NRQCD~\cite{fengGong}\\
	\\ 
	\hline
	\\
	$\langle\mathcal{O}^{\Upsilon(1S)}[{}^{3}S_1^{(1)}]\rangle$/GeV$^{3}$ & $8.39$ & $8.39$ & $8.39$ & $9.28$ \\
    	\\
	$\langle\mathcal{O}^{\Upsilon(1S)}[{}^{1}S_0^{(8)}]\rangle$/GeV$^3$ & $0.0$ & $0.0$ & $0.0$ & $0.136 \pm 0.0243$ \\	
	\\
	$\langle\mathcal{O}^{\Upsilon(1S)}[{}^{3}S_1^{(8)}]\rangle$/GeV$^3$ & $0.016 \pm 0.006$ & $0.0038 \pm 0.0019$ & $0.0029 \pm 0.0019$ & $0.0061 \pm 0.0024$ \\	
	\\
	$\langle\mathcal{O}^{\Upsilon(1S)}[{}^{3}P_0^{(8)}]\rangle$/GeV$^{5}$ & $0.07 \pm 0.03$ & $0.20 \pm 0.10$ & $0.18 \pm 0.06$ & $(-0.0093 \pm 0.005)m_b^2$ \\	
	\\
	$\langle\mathcal{O}^{\chi_{b0}(1P)}[{}^{3}P_0^{(1)}]\rangle$/GeV$^{5}$ & $2.30$ & $2.30$ & $2.30$ & $0.34$ \\
   	\\
	$\langle\mathcal{O}^{\chi_{b1}(1P)}[{}^{3}P_1^{(1)}]\rangle$/GeV$^{5}$ & $7 \pm 3$ & $11 \pm 5$ & $9 \pm 2$ & $1.02$ \\
   	\\
   	$\langle\mathcal{O}^{\chi_{b2}(1P)}[{}^{3}P_2^{(1)}]\rangle$/GeV$^{5}$ & $2.4 \pm 1.9$ & $6 \pm 4$ & $6 \pm 2$ & $1.7$ \\
   	\\
	$\langle\mathcal{O}^{\chi_{b0}(1P)}[{}^{3}S_1^{(8)}]\rangle$/GeV$^{3}$ & $0.008 \pm 0.002$ & $0.0020 \pm 0.0011$ & $0.0015 \pm 0.0012$ & $0.0094 \pm 0.0006$ \\
   	\\
	\hline
	\hline
	\end{tabular}
	\caption{The NMEs for the $\Upsilon(1S)$ and $\chi_b(1P)$ mesons as determined from our 
	fit at $p_T^{\rm cut} = 10$~GeV (the "fit A" scenario). 
	The NMEs obtained in the NLO NRQCD~\cite{fengGong} are shown for comparison.}
	\label{tab1}
	\end{table}
	
	\begin{table}[H] \footnotesize
	\centering
	\begin{tabular}{lccccc}
	\hline
	\hline
	\\
	 &  A0, fit B & JH'2013 set 1, fit B &  KMR, fit B & KMR, fit C \\
	\\ 
	\hline
	\\
	$\langle\mathcal{O}^{\Upsilon(1S)}[{}^{3}S_1^{(1)}]\rangle$/GeV$^{3}$ & $8.39$ & $8.39$ & $8.39$ & $8.39$ \\
    	\\
	$\langle\mathcal{O}^{\Upsilon(1S)}[{}^{1}S_0^{(8)}]\rangle$/GeV$^3$ & $0.0$ & $0.0$ & $0.0$ & $0.005 \pm 0.002$ \\	
	\\
	$\langle\mathcal{O}^{\Upsilon(1S)}[{}^{3}S_1^{(8)}]\rangle$/GeV$^3$ & $0.017 \pm 0.007$ & $0.004 \pm 0.004$ & $0.003 \pm 0.003$ & $0.0028 \pm 0.0017$ \\	
	\\
	$\langle\mathcal{O}^{\Upsilon(1S)}[{}^{3}P_0^{(8)}]\rangle$/GeV$^{5}$ & $0.07 \pm 0.03$ & $0.20 \pm 0.09$ & $0.18 \pm 0.05$ & $0.13 \pm 0.04$ \\	
	\\
	$\langle\mathcal{O}^{\chi_{b0}(1P)}[{}^{3}P_0^{(1)}]\rangle$/GeV$^{5}$ & $2.30$ & $2.30$ & $2.30$ & $2.30$ \\
   	\\
	$\langle\mathcal{O}^{\chi_{b1}(1P)}[{}^{3}P_1^{(1)}]\rangle$/GeV$^{5}$ & $6.4 \pm 2.1$ & $10.7 \pm 4.9$ & $9.1 \pm 2.4$ & $7.7 \pm 2.4$ \\
   	\\
   	$\langle\mathcal{O}^{\chi_{b2}(1P)}[{}^{3}P_2^{(1)}]\rangle$/GeV$^{5}$ & $3.2 \pm 1.4$ & $5.9 \pm 3.9$ & $5.5 \pm 2.1$ & $4.8 \pm 2.1$ \\
   	\\
	$\langle\mathcal{O}^{\chi_{b0}(1P)}[{}^{3}S_1^{(8)}]\rangle$/GeV$^{3}$ & $0.0$ & $0.0$ & $0.0$ & $0.0014 \pm 0.0011$ \\
   	\\
   	$\langle\mathcal{O}^{\chi_{b1}(1P)}[{}^{3}S_1^{(8)}]\rangle$/GeV$^{3}$ & $0.031 \pm 0.013$ & $0.006 \pm 0.005$ & $0.004 \pm 0.003$ & $0.0042$ \\
   	\\
   	$\langle\mathcal{O}^{\chi_{b2}(1P)}[{}^{3}S_1^{(8)}]\rangle$/GeV$^{3}$ & $0.030 \pm 0.018$ & $0.009 \pm 0.011$ & $0.008 \pm 0.009$ & $0.007$ \\
   	\\
	\hline
	\hline
	\end{tabular}
	\caption{The NMEs for the $\Upsilon(1S)$ and $\chi_b(1P)$ mesons as determined from our 
	fit at $p_T^{\rm cut} = 10$~GeV (the "fit B" scenario) and only for the KMR density at $p_T^{\rm cut} = 0$~GeV (the "fit C" scenario).}
	\label{tab1.5}
	\end{table}
	
 	\begin{table}[H] \footnotesize
	\centering
	\begin{tabular}{lcccc}
	\hline
	\hline
	\\
	7 TeV & $p_T^{\rm cut} = 10$~GeV & $p_T^{\rm cut} = 12$~GeV & $p_T^{\rm cut} = 15$~GeV & $p_T^{\rm cut} = 17$~GeV\\
	\\ 
	\hline
	\\
	A0, fit A & $0.71$ & $0.71$ & $0.72$ & $0.75$ \\	
	\\
	JH'2013 set 1, fit A & $1.35$ & $1.19$ & $1.10$ & $1.13$ \\	
	\\
	KMR, fit A & $0.77$ & $0.80$ & $0.82$ & $0.84$ \\	
	\\
	\hline
	\\
	7 + 13 TeV & $p_T^{\rm cut} = 10$~GeV & $p_T^{\rm cut} = 12$~GeV & $p_T^{\rm cut} = 15$~GeV & $p_T^{\rm cut} = 17$~GeV\\
	\\
	\hline
	\\
	A0, fit A & $2.03$ & $2.07$ & $2.12$ & $2.18$ \\	
	\\
	JH'2013 set 1, fit A & $3.68$ & $3.64$ & $3.65$ & $3.74$ \\	
	\\
	KMR, fit A & $1.72$ & $1.77$ & $1.81$ & $1.85$ \\
	\\
	\hline
	\hline
	\end{tabular}
	\caption{The dependence of $\chi^2/d.o.f.$ achieved in the 
	fit procedure (the "fit A" scenario) 
	on the choice of $p_T^{\rm cut}$ at only $\sqrt s = 7$ TeV and at $7$ and $13$~TeV combined.}
	\label{tab2}
	\end{table}
	
	\begin{table}[H] \footnotesize
	\centering
	\begin{tabular}{lccc}
	\hline
	\hline
	\\
	ATLAS+CMS & $p_T^{\rm cut} = 0$~GeV & $p_T^{\rm cut} = 5$~GeV & $p_T^{\rm cut} = 10$~GeV \\
	\\ 
	\hline
	\\
	A0, fit A & $1.89$ & $1.81$ & $1.94$ \\	
	\\
	JH'2013 set 1, fit A & $3.63$ & $3.51$ & $3.73$ \\	
	\\
	KMR, fit A & $1.47$ & $1.60$ & $1.77$ \\	
	\\
	KMR, fit C & $1.72$ & $1.89$ & $2.08$ \\	
	\\
	\hline
	\\
	Only CMS & $p_T^{\rm cut} = 0$~GeV & $p_T^{\rm cut} = 5$~GeV & $p_T^{\rm cut} = 10$~GeV \\
	\\
	\hline
	\\
	A0, fit A & $2.71$ & $2.68$ & $2.67$ \\	
	\\
	JH'2013 set 1, fit A & $5.67$ & $5.42$ & $5.30$ \\	
	\\
	KMR, fit A & $2.57$ & $2.56$ & $2.55$ \\	
	\\
	KMR, fit C & $2.95$ & $2.94$ & $2.91$ \\	
	\\
	\hline
	\hline
	\end{tabular}
	\caption{The dependence of $\chi^2/d.o.f.$ achieved in the 
	fit procedure (using NMEs from the "fit A" and "fit C" scenarios) 
	on the choice of $p_T^{\rm cut}$ for only the ATLAS data and the ATLAS and CMS data combined.}
	\label{tab3}
	\end{table}
	
\end{document}